\documentclass{aa}  

\usepackage{graphicx}
\usepackage{txfonts}
\usepackage{natbib}
\bibpunct{(}{)}{;}{a}{}{,}
\usepackage{xcolor}
\usepackage{booktabs}
\usepackage{graphicx}
\usepackage{placeins}
\usepackage{color}
\usepackage[colorlinks=true,linkcolor=black,citecolor=blue,urlcolor=blue]{hyperref}
\usepackage[noabbrev,nameinlink]{cleveref}

\usepackage[version=4]{mhchem}

\usepackage{ulem}
\def\nH{n_{\rm\langle H\rangle}}

\begin{document} 

   \title{Hydrocarbon chemistry in inner regions of planet forming disks}

    \author{J.Kanwar
          \inst{1,2,3}
          \and
          I.Kamp\inst{1}
          \and
          P.Woitke\inst{2}
          \and
          Ch.Rab\inst{4,5}
          \and 
          W.F.Thi\inst{4}
          \and
          M.Min\inst{6}
          }

   \institute{Kapteyn Astronomical Institute, University of Groningen, P.O. Box 800, 9700 AV Groningen, The Netherlands
              \email{kanwar@astro.rug.nl}
         \and
             Space Research Institute, Austrian Academy of Sciences, Schmiedlstr. 6, A-8042, Graz, Austria
        \and 
            TU Graz, Fakultät für Mathematik, Physik und Geodäsie, Petersgasse 16 8010 Graz, Austria
          \and
         Max Planck Institute for Extraterrestrial Physics, Giessenbachstrasse, 85741 Garching, Germany
         \and
         University Observatory, Faculty of Physics, Ludwig-Maximilians-Universität München, Scheinerstr. 1, 81679 Munich, Germany
         \and
         SRON Netherlands Institute for Space Research, Niels Bohrweg 4, 2333 CA Leiden, Netherlands
             }
   \date{Received 2023; accepted 2023}

  \abstract{The analysis of the mid-infrared spectra helps understanding the composition of the gas in the inner, dense and warm terrestrial planet forming region of disks around young stars. ALMA has detected hydrocarbons in the outer regions of the planet forming disk and Spitzer detected \ce{C2H2} in the inner regions. JWST- MIRI provides high spectral resolution observations of \ce{C2H2} and a suite of more complex hydrocarbons are now reported. Interpreting the fluxes observed in the spectra is challenging and radiation thermo-chemical codes are needed to properly take into account the disk structure, radiative transfer, chemistry and thermal balance. Various disk physical parameters like the gas-to-dust ratio, dust evolution including radial drift, dust growth and settling can affect the fluxes observed in the mid-IR. Still, thermo-chemical disk models were not always successful in matching all observed molecular emission bands simultaneously. 
  }
  {The goal of this project is two-fold. We analyse the warm carbon chemistry in the inner regions of the disk, i.e. within 10 au to find pathways forming \ce{C2H2} potentially missing from the existing chemical networks. Second, we analyse the effect of the new chemistry on the line fluxes of acetylene. 
  }
 {We use radiative thermo-chemical disk code {P{\small RO}D{\small I}M{\small O}} to expand the hydrocarbon chemistry that occurs in a typical standard T Tauri disks. We used the UMIST and the KIDA rate databases for collecting reactions for the species. We include a number of three-body and thermal decomposition reactions from STAND2020 network. We included isotopomers for the species that were present in the databases. The chemistry is then analysed in the regions that produce observable features in the mid-infrared spectra. The effect of expanding the hydrocarbon chemistry on the mid-infrared spectra is studied.}
 {Acetylene is formed via two pathways in the surface layers of disks: neutral-neutral and ion-neutral. They proceed via the hydrogenation of C or \ce{C+}, respectively. Thus, the abundances of C, \ce{C+}, H and \ce{H2} affect the formation of \ce{C2H2}. Therefore, also the formation of \ce{H2} indirectly affects the abundance of acetylene. Chemisorbed H is more efficient in forming \ce{H2} compared to physisorbed H at warm temperatures and hence increases the abundance of \ce{C2H2}.
}
{We provide a new extended warm chemical network that considers up to eight carbon atom long species, takes into account different isotopomers and can form the building block of PAHs \ce{C6H6}. For a standard T Tauri disk with a canonical value of gas-to-dust mass, the line fluxes increase only a factor of less than 2. JWST is now detecting hydrocarbons like methane, acetylene and  \ce{C4H2} in disks having high C/O ratio. Hence this new extended warm hydrocarbon network will aid in interpreting the observed mid-infrared fluxes.}

   \keywords{astrochemistry --
                inner disk --
                hydrocarbons -- 
                mid-IR spectra --
                JWST-MIRI
               } 

   \maketitle
%
\section{Introduction}
The inner region of a planet forming disk ($\sim$10 au) is warm (200-1000~K), dense (10$^{8}$-10$^{15}$ cm$^{-3}$) and is the nursery for the terrestrial planets \citep{Henning2013}. Understanding the rich molecular chemistry in this region can help predict the composition of terrestrial planets. Studies of the chemistry in this region are possible with mid-infrared spectroscopy which traces the warm surface layers.

Carbon is one of the most abundant elements and is essential for life. The warm molecular layer is abundant in \ce{H2}, carbon and shielded from UV photons by dust, \ce{H2} \citep{Draine1996}, \ce{H2O} \citep{Bethell2009, Duval2022...934L..25D}, thus paving the way for a rich organic chemistry. Carbon based molecules like CO, \ce{CO2}, C$_2$H$_2$ have been observed using Spitzer \citep[e.g.][]{Carr2011,Salyk2008,Pontoppidan2010}. Other hydrocarbons like c-C$_3$H$_2$ \citep{Qi2013a}, C$_2$H \citep{Bergin2016} have been detected in the outer disks by the MAPS consortium \citep{JD2021A,viv2021}. With the advent of JWST, a plethora of new species are discovered due to its high sensitivity and increased spectral resolution compared to Spitzer. Among the new detections \citep[][\citealt{Arabhavi2023}]{Tabone2023} are various hydrocarbons emitting in the mid-infrared region as predicted by \cite{Bast2013}.

There have been various studies on hydrocarbon chemistry in different environments (temperatures, pressures initial abundances, radiation etc.) like in molecular clouds, AGB stars, planetary nebulae and planet forming disks. Carbon chains can form at the early stages of cloud evolution before C gets locked in CO in warm carbon chain chemistry in molecular clouds. Experimental study by \cite{Santoro2020} investigated the gas-phase interaction of C and \ce{C2} with \ce{C2H2} leading to polyacetylenic chains and poly-cyclic aromatic hydrocarbons (PAHs) for the outer layers of C-rich AGB and protoplanetary nebulae. Benzene is shown to form in a bottom up approach in the inner 3\,au of the planet forming disk and its abundance structure is affected by the uncertainty in the adsorption energies assumed for it \citep{Woods2007}.
\cite{Kress2010} studied the formation of poly-cyclic aromatic hydrocarbons (PAHs) and calculated the position of the soot line beyond which PAHs are destroyed in disks. A comparative study on chemistry around the stars with spectral types from A to M showed different abundance structures of \ce{C2H2} and set of pathways leading to its formation in X-ray and UV dominated regions \citep{Walsh2015}. 

Hydrocarbons have been detected in the ISM and molecular clouds with only \ce{C2H2} detected in disks around T Tauri stars and \ce{CH4}, \ce{C4H2}, \ce{C6H6} detected \citep{Tabone2023} in disks around very low mass stars in the mid-IR. There have been numerous attempts to reproduce the observational flux levels through thermo-chemical modelling. Earlier work \citep[e.g.][]{Woitke2018,Greenwood2019, Greenwood2019a, Anderson2021} proposed high gas-dust, dust evolution (including radial drift, growth, settling) and a high C/O ratio to interpret the observed fluxes of \ce{C2H2} and \ce{H2O}. Little work went in looking in more detail into the formation pathways of \ce{C2H2} which explains the need to revisit the chemical networks used widely in the disk community.

Hence, in this paper, the focus is on expanding the hydrocarbon chemistry and studying how this affects the mid-infrared spectra especially in the light of new JWST data. The radiation thermo-chemical modelling code {P{\small RO}D{\small I}M{\small O}} \citep{Woitke2009} is used to determine the disk structure of a typical T Tauri disk \citep{Woitke2016}. The hydrocarbon chemistry is expanded beyond the large DIANA chemistry \citep{Kamp2017}. 
The key ingredients of the disk modelling code {P{\small RO}D{\small I}M{\small O}}, the large DIANA chemical network and the description of the extension of the chemical network for hydrocarbons is provided in section \ref{Modeling}. Section~\ref{investigating the hydrocarbon chemistry} identifies pathways of \ce{C2H2} formation, presents network diagrams and analyses the hydrocarbon chemistry in a typical T Tauri disk and highlights the differences between using the UMIST \citep{McElroy2013} and the KIDA \citep{Wakelam2012}\footnote{Note that we use here the published network file kida.uva.2014 from \citet{wakelam2015}, referred to as KIDA2014 in the remainder of this work.} chemical rate databases. The implications of the new chemical network on the mid infrared spectra is described in section \ref{Mid-IR spectra}. Section~\ref{Discussion} discusses the results followed by conclusions.

\section{Modeling}\label{Modeling}
\subsection{Physical disk Modeling}\label{Physical Modeling}

{P{\small RO}D{\small I}M{\small O}} (Version:\,2.0-421754a9) is a radiation thermo-chemical code that models the physical and chemical structure of planet forming disks \citep{Woitke2009,Woitke2016}. It assumes an axis-symmetric, Keplerian, irradiated disk. It performs 3D ray based dust continuum radiative transfer to obtain the dust temperature and the wavelength dependent radiation field in the disk \citep{Woitke2009}. 

\begin{figure}[h!]
    \resizebox{\hsize}{!}{\includegraphics{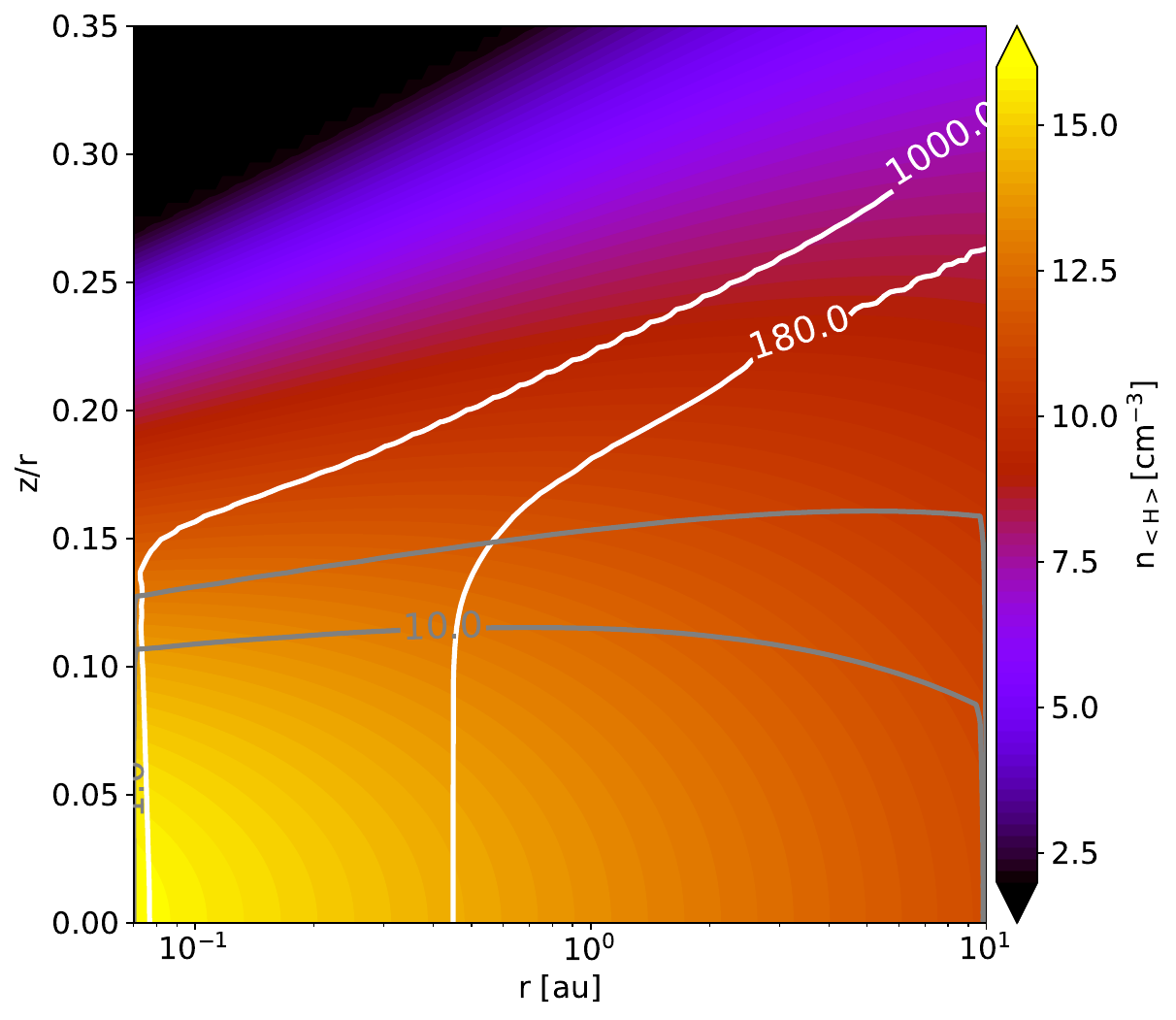}}
    \caption{The particle density ($\rm n_{<H>}$) of the disk in all the subsequent models. The white lines corresponds to the gas temperature of 180\,K and 1000\,K. The gray contours depict the combined radial and vertical $A_{\rm V}=1$ and $A_{\rm V}$=10.}
    \label{struc.pdf}
\end{figure}

\begin{figure}[h]
    \hspace*{-2mm}
    \resizebox{\hsize}{!}{\includegraphics{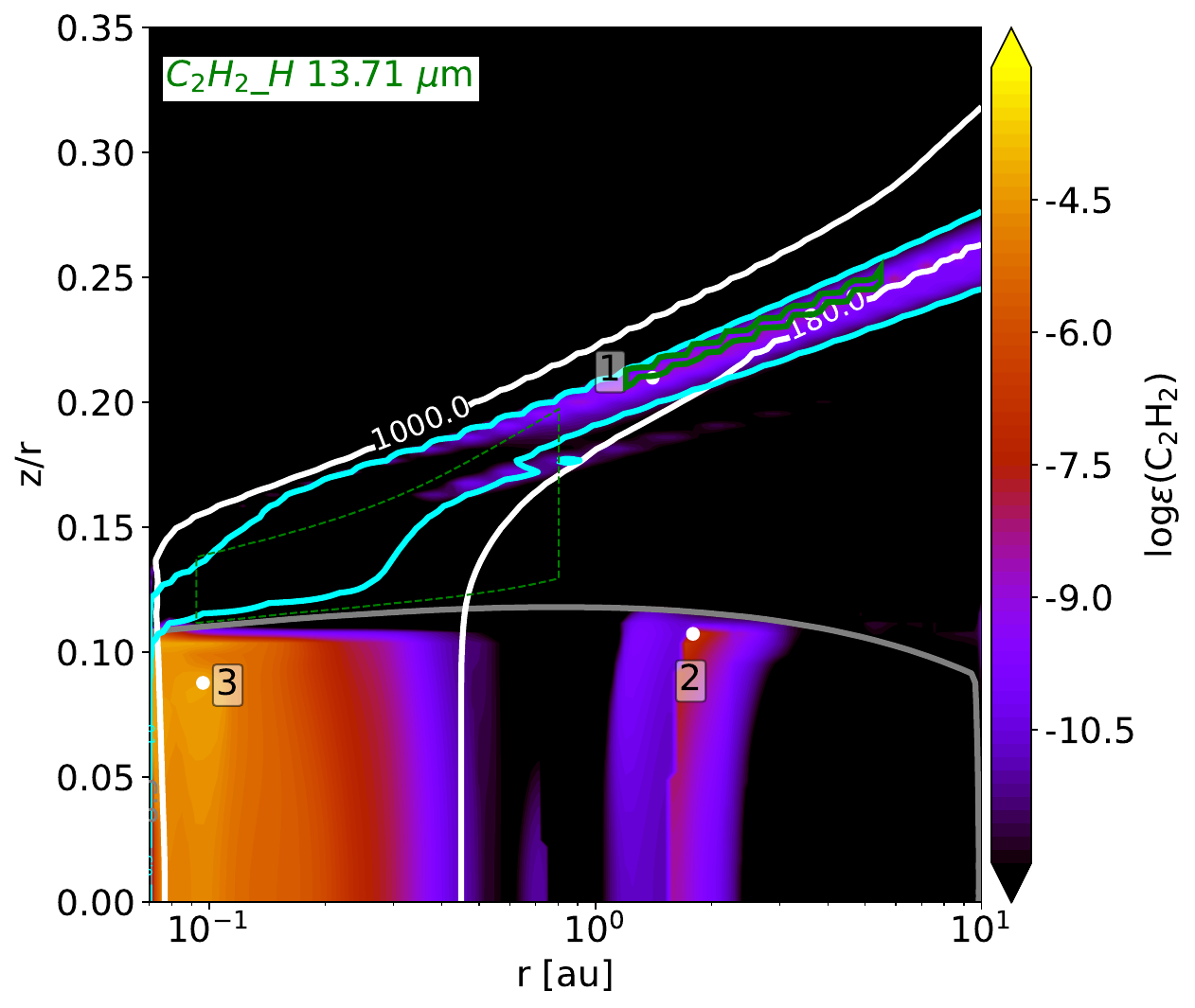}}
    \caption{The abundance of C$_2$H$_2$ with respect to $\nH$ obtained from the fiducial disk model. The white lines show the gas temperature contours for 180\,K and 1000\,K The gray contour depicts $A_{\rm V}=8.5$. The green dotted line shows the region from which 50\% of the continuum emits at $\sim$13.7$ \rm \,\mu m$ (representative of the $ \rm \nu_5$ fundamental band). The green solid line shows the region where 50\% of the flux is emitted by C$_2$H$_2$ at 13.7$ \rm \,\mu m$. The white dots indicate the grid points for which the chemistry is subsequently analysed in detail. The cyan contours depict where 20 and 99\% of hydrogen is in the form of \ce{H2}.} 
    \label{fig_63}
\end{figure}

We model a steady-state standard T Tauri star-disk system with a grid point resolution of 150x100 cells. Our disk parameters are the same as in \cite{Woitke2018} except for the radial extent $R_{\rm out}$. The emitting region of molecules like C$_2$H$_2$ in the mid-IR are limited to within 10~au \citep{Woitke2018}. Hence, we cut the radial extent of the disk to 10~au to be able to increase the spatial resolution for the inner regions. Our corresponding disk mass $M_{\rm disk}$ inside 10~au is only 9.495$\cdot$10$^{-4}$~M$_{\odot}$. We also use the canonical gas-to-dust mass ratio of 100. Figure\,\ref{struc.pdf} shows the physical structure of the disk model.

\subsection{Thermo-Chemical disk Modeling}
\begin{table}[]
\caption{The list of all the elements along with their abundances used in all the models.}
\label{elements}
\begin{tabular}{@{}lll@{}}
\hline
Elements & 12 + log$\epsilon$ & Abundance (relative to H) \\ \hline
H & 12 & 1 \\
He & 10.98 & 9.64$\cdot$10$^{-2}$ \\
C & 8.14 & 1.38$\cdot$10$^{-4}$ \\
N & 7.90 & 7.94$\cdot$10$^{-5}$ \\
O & 8.48 & 3.02$\cdot$10$^{-4}$ \\
Ne & 7.95 & 8.91$\cdot$10$^{-5}$ \\
Na & 3.36 & 2.29$\cdot$10$^{-9}$ \\
Mg & 4.03 & 1.07$\cdot$10$^{-8}$ \\
Si & 4.24 & 1.74$\cdot$10$^{-8}$ \\
S & 5.27 & 1.86$\cdot$10$^{-7}$ \\
Ar & 6.08 & 1.20$\cdot$10$^{-6}$ \\
Fe & 3.24 & 1.74$\cdot$10$^{-9}$ \\
PAH & 5.48 & 3.019$\cdot$10$^{-7}$\\
\hline
\end{tabular}%
\end{table}
The gas temperature in {P{\small RO}D{\small I}M{\small O}\;}is determined by the balance between the heating and cooling processes as listed in \cite{Woitke2009}. 
The code uses the kinetic rate approach to obtain the chemical composition in the disk. Additional heating/cooling processes namely the line cooling by molecules emitting in the mid-IR like H$_2$O, \ce{C2H2}, CH$_4$, CO$_2$, NH$_3$, HCN, OH is included in the model \citep{Woitke2018}. 

\begin{figure*}[h!]
    \centering
    \includegraphics[width=\linewidth]{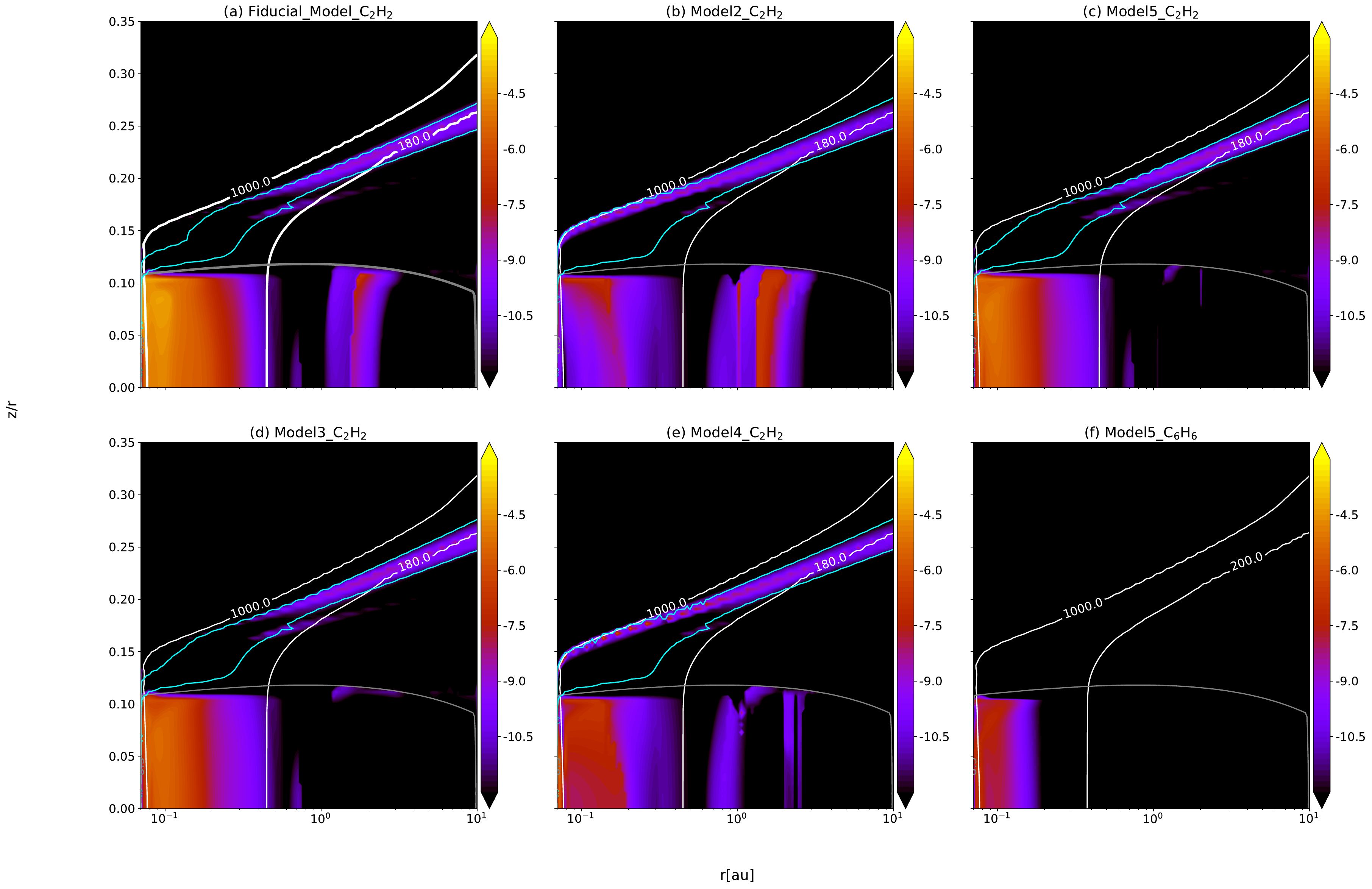}
    \caption{The abundance of species C$_2$H$_2$, \ce{C6H6} in different models are shown. The gray contour corresponds to A$_V$=8.5. The temperature contours of 1000~K and 180~K are shown in white. The cyan contours represent where 20 and 99.8\% of hydrogen is in the form of \ce{H2}.}
    \label{5_plots}
\end{figure*}

\begin{figure}[h!]
    \centering
    \includegraphics[width=\linewidth]{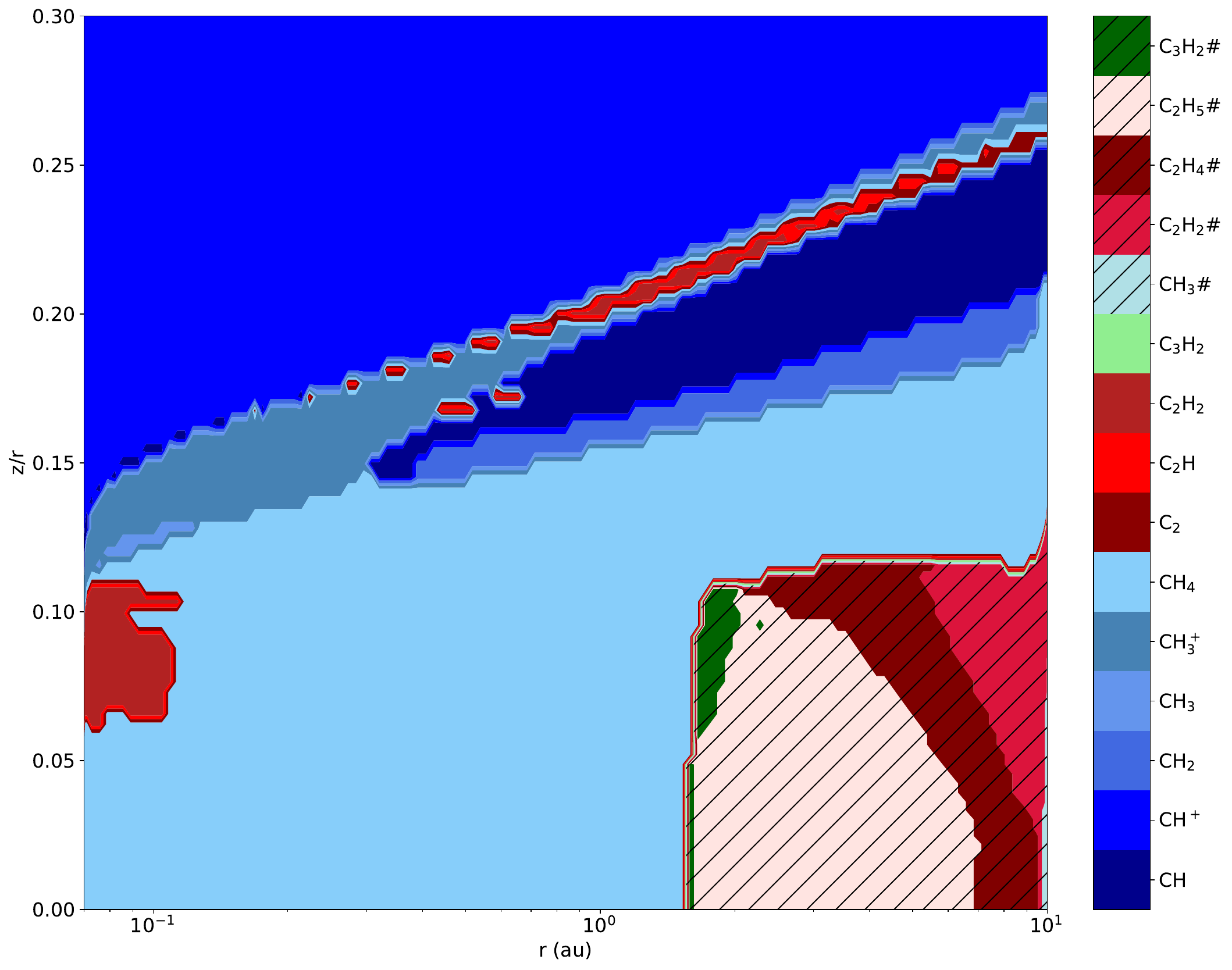}
    \caption{The distribution of the most abundant hydrocarbons in the fiducial disk model (large DIANA chemistry and UMIST2012 rate coefficients). Each grid point depicts the most abundant hydrocarbon at that location. We do not consider C or \ce{C+} when finding the abundant hydrocarbons.}
    \label{HC_M}
\end{figure}
\begin{figure}
    \includegraphics[width=\linewidth]{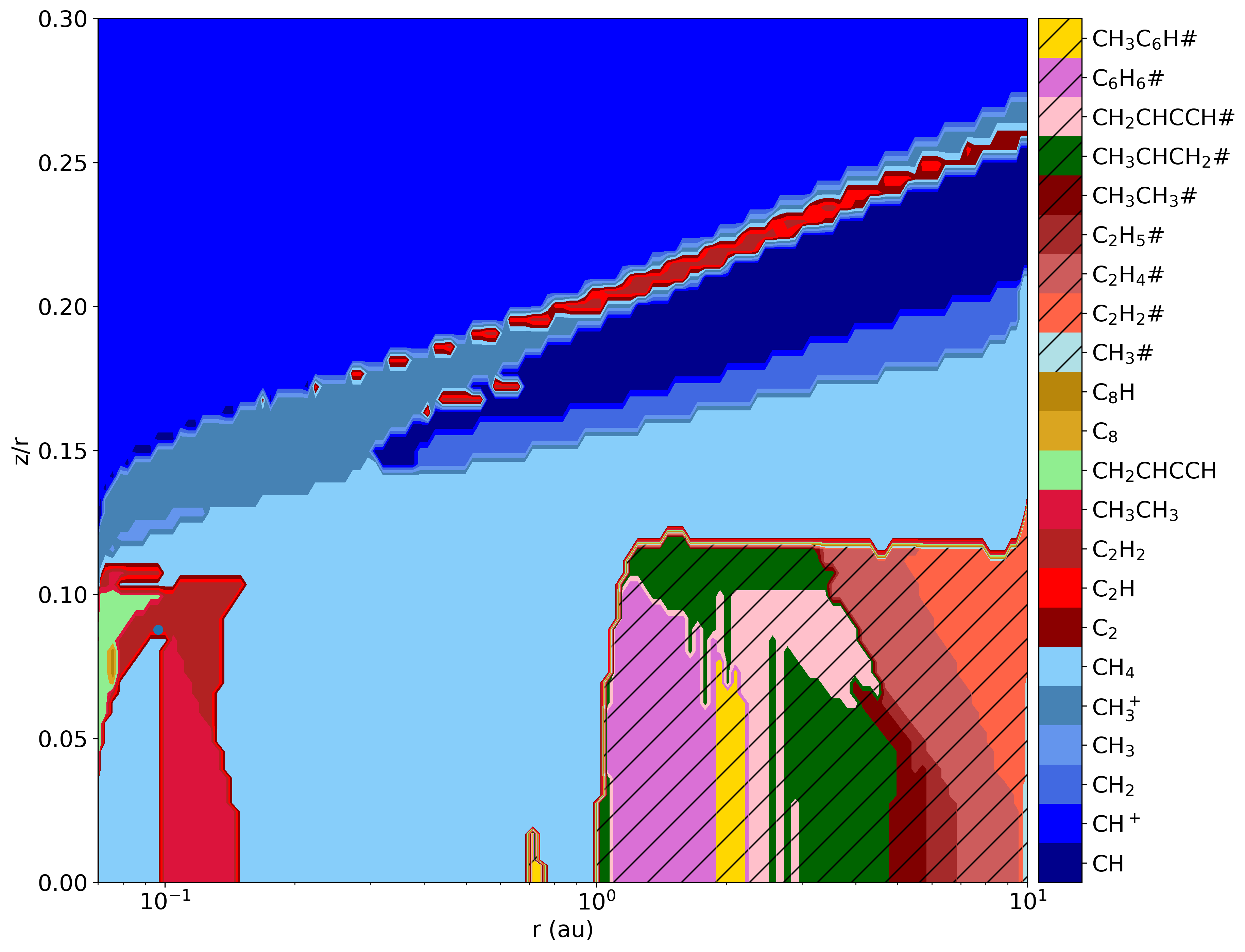}
    \caption{The distribution of the most abundant hydrocarbons in model 5 (final extended chemical network).}
    \label{HC_M1}
\end{figure}

The chemical network includes the gas-phase chemistry and gas-grain desorption (thermal desorption, photo-desorption and cosmic-ray desorption) and freeze-out. The chemistry is modeled using 13 elements, namely H, He, C, N, O, Ne, Na, Mg, Si, S, Ar, Fe and PAH. Table\,\ref{elements} lists the elemental abundances used in all the models \citep{Woitke2016}; we use C/O ratio of 0.457. The fiducial model uses the large DIANA chemical network consisting of 235 species, including atoms, molecules, corresponding positive ions, protonated ions and ices \citep{Kamp2017}. 
There are 3037 reactions in the fiducial model that largely originate from the UMIST2012 database \citep{McElroy2013}. We add two reactions from the KIDA 2014 database, which are explained in Sect.~ \ref{Building the Chemical network}. The network has a number of specific additional reactions as mentioned in \cite{Kamp2017} : X-ray chemistry \citep{Meijerink2012,Crab2018}, PAH chemistry (C$_{54}$H$_{18}$ is taken as a representative PAH) including adsorption and freeze out \citep[see references in][]{Kamp2017}, collider reactions from the UMIST 2006 database and photodissociation reactions for the molecular ions \citep{Heays2017}. Photorates are calculated using the local wavelength dependent radiation field inside the disk \citep{Kamp2010}. We include a number of three body and thermal decomposition reactions taken from STAND2020 \citep{Rimmer2016} listed in the Appendix\,\ref{3body reactions}. We take only those reactions from STAND2020 which have valid data for the forward direction \footnote{A+B+M $\rightarrow$ C+D+M, reverse of such a reaction is not added}; the backward reactions for the same are not included. H$_2$ formation is included following \cite{Cazaux2004}. We explore the impact of using a different \ce{H2} formation rate \citep{Cazaux2010} in Sect.~ \ref{Effect of}.

\subsection{Expanding the network for longer hydrocarbon molecules} \label{Building the Chemical network}

\begin{table*}[]
\caption{The list of all the hydrocarbons present in the extended chemical network.}
\begin{tabular}{@{}cccccccc@{}}
\toprule
C & 2C & 3C & 4C & 5C & 6C & 7C & 8C \\ \midrule
\textcolor{gray}{CH} & \textcolor{gray}{\textbf{C$_2$}} & \textcolor{gray}{\textbf{C$_3$}} & \textcolor{gray}{\textbf{C$_4$}} & \textbf{C$_5$} & \textbf{C$_6$} & \textbf{C$_7$} & \textbf{C$_8$} \\
\textcolor{gray}{CH$_2$} & \textcolor{gray}{C$_2$H} & \textcolor{gray}{C$_3^+$} & \textcolor{gray}{C$_4^+$} & C$_5^+$ & C$_6^+$ & C$_7^+$ & C$_8^+$ \\
\textcolor{gray}{CH$_3$} & \textcolor{gray}{\textbf{C$_2$H$_2$}} & \textcolor{gray}{C$_3$H} & \textcolor{gray}{C$_4$H$^+$} & C$_5$H & C$_6$H & C$_7$H & C$_8$H \\
\textcolor{gray}{\textbf{CH$_4$}} & \textcolor{gray}{C$_2$H$_3$} & \textcolor{gray}{C$_3$H$^+$} & C$_4$H & C$_5$H$^+$ & C$_6$H$^+$ & C$_7$H$^+$ & C$_8$H$^+$ \\
\textcolor{gray}{CH$^+$} & \textcolor{gray}{\textbf{C$_2$H$_4$}} & \textcolor{gray}{\textbf{C$_3$H$_2$}} & \textbf{HC$_4$H} & \textbf{C$_5$H$_2$} & \textbf{C$_6$H$_2$} & \textbf{C$_7$H$_2$} & \textbf{C$_8$H$_2$} \\
\textcolor{gray}{CH$_2^+$} & \textcolor{gray}{C$_2$H$_5$} & \textcolor{gray}{C$_3$H$_2^+$} & C$_4$H$_3$ & C$_5$H$_2^+$ & C$_6$H$_2^+$ & C$_7$H$_2^+$ & C$_8$H$_2^+$ \\
\textcolor{gray}{CH$_3^+$} & \textcolor{gray}{C$_2^+$} & \textcolor{gray}{C$_3$H$_3^+$} & \textbf{CH$_2$CHCCH} & C$_5$H$_3^+$ & C$_6$H$_3^+$ & C$_7$H$_3^+$ & C$_8$H$_3^+$ \\
\textcolor{gray}{CH$_4^+$} & \textcolor{gray}{C$_2$H$^+$} & \underline{CCCH (l-C$_3$H)} & \textbf{CH$_2$CHCHCH$_2$} & \textbf{CH$_3$C$_4$H} & C$_6$H$_4^+$ & C$_7$H$_4^+$ & C$_8$H$_4^+$ \\
\textcolor{gray}{CH$_5^+$} & \textcolor{gray}{C$_2$H$2^+$} & H$_2$CCC (l-C$_3$H$_2$) & C$_4$H$_2^+$ & CH$_3$C$_4$H$^+$ & C$_6$H$_5^+$ & C$_7$H$_5^+$ & C$_8$H$_5^+$ \\
\textcolor{gray}{CH$\#$} & \textcolor{gray}{C$_2$H$_3^+$} & CH$_2$CCH & C$_4$H$_3^+$ & C$_5$H$_5^+$ & \textbf{C$_6$H$_6$} & \textbf{CH$_3$C$_6$H} & C$_8\#$ \\
\textcolor{gray}{CH$_2\#$} & \textcolor{gray}{C$_2$H$_4^+$} & \textbf{CH$_3$CCH} & C$_4$H$_4^+$ & C$_5\#$ & \underline{C$_6$H$_6^+$} & C$_7\#$ & C$_8$H$\#$ \\
\textcolor{gray}{CH$_3\#$} & \textcolor{gray}{C$_2$H$_5^+$} & \underline{\textbf{CH$_2$CCH$_2$}} & C$_4$H$_5^+$ & C$_5$H$\#$ & C$_6$H$_7^+$ & C$_7$H$\#$ & C$_8$H$_2\#$ \\
\textcolor{gray}{CH$_4\#$} & \textbf{CH$_3$CH$_3$} & \textbf{CH$_3$CHCH$_2$} & C$_4$H$_7^+$ & C$_5$H$_2\#$ & C$_6\#$ & C$_7$H$_2\#$ &  \\
 & CH$_3$CH$_3^+$ & CH$_2$CCH$^+$ & \textcolor{gray}{C$_4\#$} & CH$_3$C$_4$H$\#$ & C$_6$H$\#$ & CH$_3$C$_6$H$\#$ &  \\
 & C$_2$H$_7^+$ & C$_3$H$_4^+$ & C$_4$H$\#$ &  & C$_6$H$_2\#$ &  &  \\
 & \textcolor{gray}{C$_2\#$} & C$_3$H$_5^+$ & HC$_4$H$\#$ &  & C$_6$H$_6\#$ &  &  \\
 & \textcolor{gray}{C$_2$H$\#$} & \underline{C$_3$H$_6^+$} & C$_4$H$_3\#$ &  &  &  &  \\
 & \textcolor{gray}{C$_2$H$_2\#$} & \underline{C$_3$H$_7^{+}$} & CH$_2$CHCCH$\#$ &  &  &  &  \\
 & \textcolor{gray}{C$_2$H$_3\#$} & \textcolor{gray}{C$_3\#$} & CH$_2$CHCHCH$_2\#$ &  &  &  &  \\
 & \textcolor{gray}{C$_2$H$_4\#$} & \textcolor{gray}{C$_3$H$\#$} &  &  &  &  &  \\
 & \textcolor{gray}{C$_2$H$_5\#$} & \textcolor{gray}{C$_3$H$_2\#$} &  &  &  &  &  \\
 & CH$_3$CH$_3\#$ & \underline{CCCH$\#$} &  &  &  &  &  \\
 &  & H$_2$CCC$\#$ &  &  &  &  &  \\
 &  & CH$_2$CCH$\#$ &  &  &  &  &  \\
 &  & CH$_3$CCH$\#$ &  &  &  &  &  \\
 &  & \underline{CH$_2$CCH$_2\#$} &  &  &  &  &  \\
 &  & CH$_3$CHCH$_2\#$ &  &  &  &  &  \\ \bottomrule
\end{tabular}
\label{Table3}
\tablefoot{The species in gray were present in the large DIANA chemical network (and are present in all models). The species in black are added to this network. All these species are present in model 5. All the species except the underlined ones are present in models 3 and 4. The stable species are marked in bold. The ices of neutral species are represented by \#.}
\end{table*}
We aim to analyse the chemical formation and destruction pathways of \ce{C2H2} and quantify the effect of chemistry on the \ce{C2H2} abundance and its mid-IR spectra. Our goal is to find potentially missing chemical pathways from the existing chemical network and study their impact on the disk. To do so, we first expand the large DIANA chemical network by adding additional longer hydrocarbon species.

We decided to restrict ourselves to the chemistry up to the most simple cyclic aromatic hydrocarbon, namely \ce{C6H6}. We included species up till eight carbon atoms to take into account the destruction of larger hydrocarbons to form the stable C$_6$H$_6$ molecule. The negative ions are omitted except for H$^-$. 
The rule formulated in \citet{Kamp2017} which states to include the ions and protonated forms of the neutral stable closed shell species is followed. We identify the stable species as C$_2$H$_6$, C$_3$H$_4$, C$_3$H$_6$, C$_4$H$_2$, C$_4$H$_4$, C$_4$H$_6$, C$_5$, C$_5$H$_2$, C$_5$H$_4$, C$_6$, C$_6$H$_2$, C$_6$H$_6$, C$_7$, C$_7$H$_2$, C$_7$H$_4$, C$_8$, C$_8$H$_2$. Their ions and protonated forms are added in the network except for C$_4$H$_6$. Only the protonated form of this species is added in the network because destruction reactions for \ce{C4H6+} were unavailable in the rate databases. 

In a first step, the hydrocarbon species and their isotopomers, if any, for which the destruction and formation reactions are available in the UMIST2012 database are included in the network. The network has the cyclic and linear isotopomers of \ce{C3H} denoted as \ce{C3H} and \ce{CCCH}, C$_3$H$_2$ denoted as \ce{C3H2} and \ce{H2CCC}, 
C$_3$H$_3^+$ denoted as \ce{C3H3+} and \ce{CH2CC+}, respectively. It also has both linear isotopomers of C$_3$H$_4$ denoted as CH$_2$CCH$_2$ and CH$_3$CCH.

The reactions for \ce{CCCH} are taken from KIDA2014. 
Both the KIDA2014 and the UMIST2012 database do not provide the reactions for the cyclic counterpart \ce{c-C3H+} which might be because this species has never been detected in space.
The network also misses C$_6$H$_5$ and C$_6$H$_4$ as neither of the two databases provides the destruction and formation reactions for these species.
The KIDA2014 database misses C$_3$H$_6^+$, C$_3$H$_7^+$, C$_6$H$_6^+$, CH$_2$CCH$_2$ and UMIST misses CCCH.

This resulted in a total of 92 new hydrocarbon species added to the large chemical DIANA network. Rates for their reactions are takes from either UMIST2012 or KIDA2014. Table~\ref{Table3} summarizes this list and shows the stable species in bold font. Neutral hydrocarbons including the radicals can freeze on the dust grains. The adsorption energies for these neutral species are taken also from the UMIST2012 database. We assumed the same adsorption energies for both the isomers of \ce{C3H4} due to the lack of individual data. The enthalpy of formation at 0\,K is taken from the UMIST and KIDA database, respectively. We were not able to find the heat of formation for C$_3$H$_6^+$, C$_4$H$_4^+$, C$_4$H$_7^+$, C$_5$H$^+$, C$_5$H$_2^+$, C$_5$H$_3^+$, C$_5$H$_4$, C$_6^+$, C$_6$H$^+$,C$_6$H$_3^+$,C$_6$H$_6^+$, C$_6$H$_7^+$, C$_7$H$^+$, C$_7$H$_2^+$, C$_7$H$_3^+$, C$_7$H$_4^+$, C$_7$H$_5^+$, CH$_3$C$_6$H, C$_8^+$, C$_8$H$^+$, C$_8$H$_3^+$, C$_8$H$_4^+$, C$_8$H$_5^+$ in the UMIST, the KIDA or the NIST database. Similar to the adsorption energies, we assumed then the enthalpy of formation for isotopomers as equal for these species.
\begin{table}[]
\caption{Species missing in the databases. Stable species are marked in bold.}
\begin{tabular}{ll} 
\hline
\hline
Only in UMIST2012 & Only in KIDA2014 \\ \hline
\ce{C3H6+} &  CCCH\\
C$_3$H$_7^+$  &  \\
C$_6$H$_6^+$  &        \\
\textbf{CH$_2$CCH$_2$} &        \\
\hline
\end{tabular}
\label{tab:my-table}
\end{table}

\subsection{Chemical Networks}
We aim to investigate the changes in the chemistry of \ce{C2H2} from using different sets of rate coefficients from UMIST2012 and KIDA2014 after expanding the chemical network. We can then analyse the impact of the extended chemical network on the mid-IR spectra from the disks. To build up our understanding, a series of models was used. 

Our fiducial model uses the large DIANA chemical network with 235 species \citep{Kamp2017}, plus the set of three body and thermal decomposition reactions listed in Appendix\,\ref{3body reactions}. The heating and cooling processes that determine the gas temperature are intertwined with the chemistry. Hence, we decided to fix the physical structure of the disk (gas, dust temperatures and densities) to that of the fiducial model for all subsequent models. Only the chemical concentrations were recalculated. This allows to isolate the effect of chemical rates on the species abundances, line emitting regions of the molecules and the mid-IR line fluxes. 

Each model has a barrier-less charge exchange and a dissociative recombination formation reaction for C$_2$H$_2$ taken from the KIDA2014 database:
\begin{equation} \label{1}
    \ce{C} + \ce{C2H2+} \rightarrow \ce{C2H2} + \ce{C+} 
\end{equation}
\begin{equation} \label{2}
    \ce{C3H3+} + \ce{e-} \rightarrow \ce{CH} + \ce{C2H2} .
\end{equation}
\footnote{all the reactions referenced are from within the text}These reactions constitute a dominant formation pathway of \ce{C2H2} when using KIDA2014 and are therefore added in all the models.

Table \ref{Table 2} summarises the different models that we use in this work. The chemical database and the size of the chemical network are varied in the models. Model 2 uses the large DIANA chemical network with the rates calculated using the rate coefficients provided by the KIDA2014 database. Model 3 and model 4 use the new extended chemical network consisting of all the species in Table\,\ref{Table3} except the underlined ones and use rate coefficients for reactions primarily taken from UMIST2012 and KIDA2014, respectively. They have the same species to facilitate the comparison between the two databases. This is why they lack seven species, namely C$_3$H$_6^+$, C$_3$H$_7^+$, C$_6$H$_6^+$, CH$_2$CCH$_2$(C$_3$H$_4$) and its corresponding ice CH$_2$CCH$_2$\#, and CCCH and its corresponding ice CCCH\# \footnote{\# represents ices of the corresponding species}. The databases are not complete in providing all the required reactions and species. For example, the UMIST2012 database clubbed \ce{c-C3H2+} and \ce{l-C3H2+} together as \ce{C3H2+} which is used in model 3. We therefore use \ce{c-C3H2+} from the KIDA2014 database as its counterpart in model 4. \\
Model 5 has the missing seven species included and uses the UMIST2012 database and a few selected reactions added from the KIDA2014 database for CCCH. Model 5 thus presents the complete extended chemical network.
Model 5 also includes the following dissociative recombination reaction with electrons  
\begin{equation} \label{3}
    \ce{C4H3+} + \ce{e-} \rightarrow \ce{C2H} + \ce{C2H2} .
\end{equation}
The rate coefficient is taken from the KIDA2014 database.

\begin{table}
\caption{List of models used in this paper specifying the details of the chemical network.}
\begin{tabular}{lllll}
\hline
\hline
Model \# & database & \# of species &  \# of reactions&  \\ \hline
Fiducial & UMIST2012    & 235     &  3036&  \\ 
Model 2  & KIDA2014     & 235     &  3072&  \\ 
Model 3  & UMIST2012    & 320     &  4004&  \\ 
Model 4  & KIDA2014     & 320     &  4150&  \\ 
Model 5  & UMIST2012    & 327     &  4121 &  \\ \hline
\end{tabular}
\label{Table 2}
\end{table}

\section{The hydrocarbon chemistry}\label{investigating the hydrocarbon chemistry}
\begin{figure*}[h!]
    \centering
    \includegraphics[width=.9\linewidth]{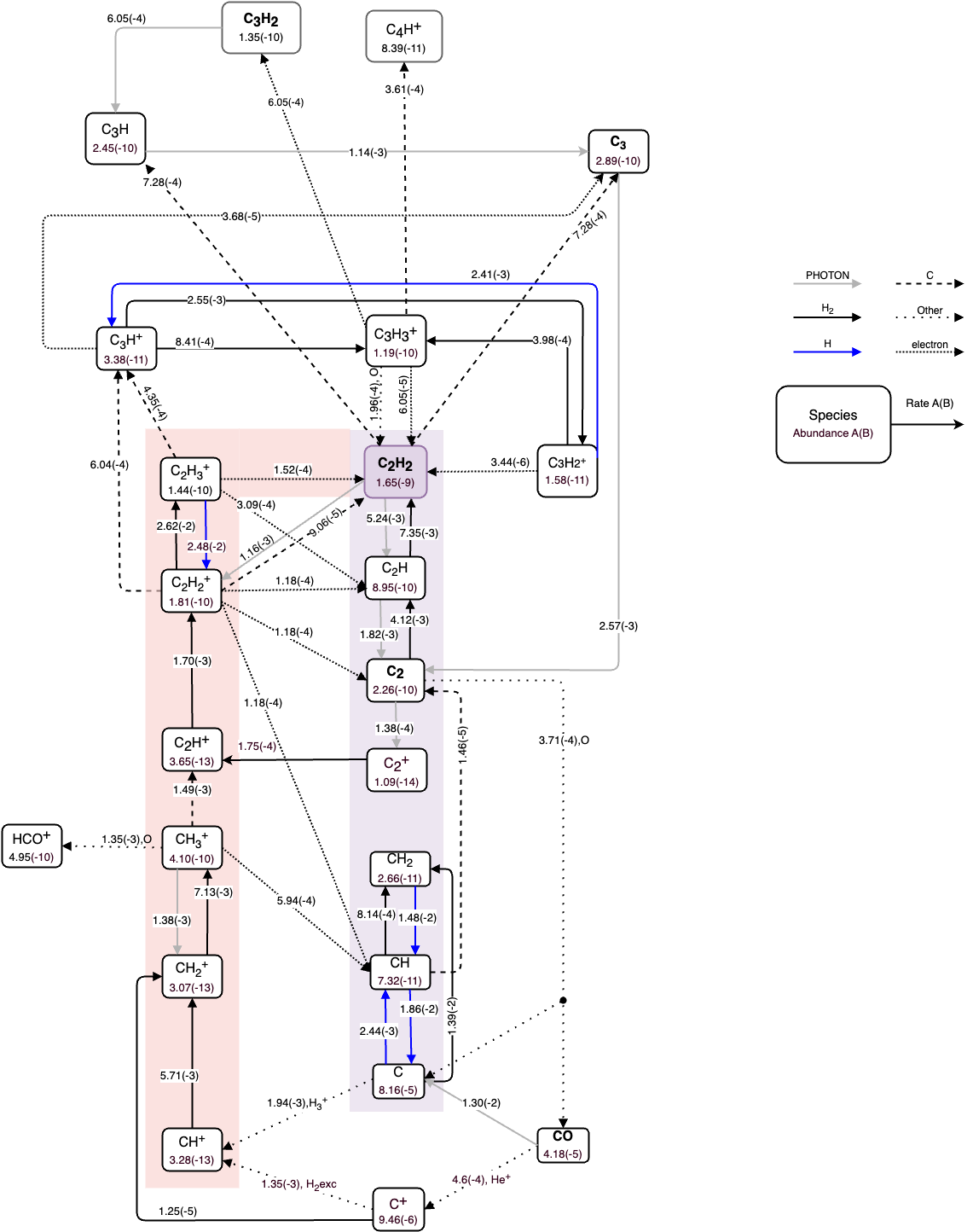}
    \caption{The chemical network centered around C$_2$H$_2$ depicting the dominant formation and destruction pathways of parent molecules for the fiducial model in the surface layers (grid point 1). It includes large DIANA chemical species and uses the UMIST2012 rate coefficients. The rates are indicated on the paths. The A(B) notation represents A$\cdot$10$^B$. The species abundance in the surface layer in the fiducial model for a grid point is noted below the species. 
}
    \label{UMIST+235}
\end{figure*}

\begin{figure*}[h]
    \centering
    \includegraphics[width=.8\linewidth]{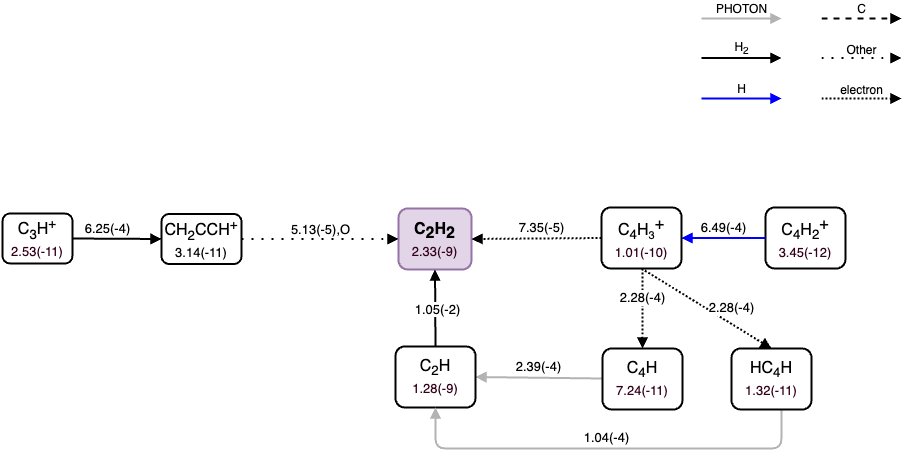}
    \caption{Zoomed in chemical network showing the formation pathways that become active after adding the longer hydrocarbons. The abundances and rates correspond to model~3 (extended hydrocarbon network using the UMIST2012 rate database). This also highlights the major differences in the formation pathways of \ce{C2H2} between Fig\,\ref{UMIST+235} and \ref{327_network}.}
    \label{UMIST+317}
\end{figure*}

\begin{table*}[]
\caption{Properties of the grid point for which the chemistry was analysed in various models.} 
\begin{tabular}{cccccccccc} \hline
point & r,z (au)   & T$_{\rm gas}$ [K] & T$_{\rm dust}$ [K] & A$_v^{\rm rad}$ & A$_v^{\rm ver}$ & n$_{\rm <H>}$ [cm$^{-3}$] & $\chi$ & remark & reservoirs in fiducial model \\ \hline
1& 1.41, 0.30  &300          &230           &1.10       &0.021      &$5.6\cdot10^9$ & $3.3\cdot10^5$ & surface layer & \ce{H2},\ce{He},\ce{O},\ce{C},\ce{CO}\\
2& 1.79, 0.19  &85           &85            &3200       &13         &$1.3\cdot10^{12}$ & 0  & outer midplane & \ce{H2},\ce{He},\ce{CO},\ce{C3H2}\# \\
3& 0.1, 0.0085 &480          &480           &1700       &62        &$2.2\cdot10^{14}$ & 0 & inner midplane & \ce{H2},\ce{He},\ce{CH4},\ce{H2O},\\ \hline            
\end{tabular}\\[1mm]
\label{property}
\tablefoot{Location describes the radial and vertical direction in the disk model. n$_{\rm <H>}$ is the particle density of H, A$_{\rm v,rad}$ and A$_{\rm v,ver}$ are the radial and vertical visual extinction, $\chi$ is the UV radiation field at that location in units of drain field. The final column mentions the main form of C, H, \ce{He}, O in the reservoir in the surface layers in the fiducial model.} 
\end{table*}

The abundance of \ce{C2H2} in the fiducial model is shown in Fig~\ref{fig_63}. \ce{C2H2} resides in a thin surface layer and two larger reservoirs in the midplane around 0.1\,au and 2\,au. This general qualitative structure of the \ce{C2H2} abundance is the same in all the models no matter which size of network/database we use; only the radial extent changes slightly. 

To understand the formation of \ce{C2H2} in our disk model, we have chosen three representative grid points from the three reservoirs shown as white circles in Fig.~\ref{fig_63}. Grid point 1 is the representative point for surface chemistry that lies around the emitting region of the C$_2$H$_2$ transition at 13.706~$\mu$m above the continuum emitting surface at 13.7~$\mu$m. This particular transition is chosen as it has the highest flux in the fiducial model. Grid points 2 and 3 reside in regions of maximum abundance of \ce{C2H2} in the two reservoirs in the midplane. Table~\ref{property} summarises the properties of the locations that are analysed. The typical abundance in the models in the surface layers (grid point~1) and the regions beyond $r\,\leq\,1$~au (grid point 3) is of the order of 10$^{-9}$ and 10$^{-5}$, respectively. The surface layer is confined to a region where the abundance of \ce{H2} is higher than $\sim$0.05, the abundance of H is higher than $\sim$0.0019 and the abundance of C is of the order of $\sim$ 10$^{-6}$. The two regions in the midplane are confined by the absence of electrons. In regions with an electron abundance $\geq\,10^{-11.5}$, the abundance of \ce{C2H2} is low ($\leq$10$^{-12}$). In the second reservoir around 2 au O is depleted from the gas as it forms \ce{H2O} ice. The gas then becomes C-rich and that gives rise to carbon chemistry.

First, the chemistry of \ce{C2H2} at grid point 1 in the surface layer is analysed. The major formation and destruction pathways of \ce{C2H2} in the fiducial model are shown in Fig~\ref{UMIST+235}. Carbon, being the key ingredient of any hydrocarbon chemistry is unlocked by the photodissociation of CO by UV photons. It can also be unlocked as \ce{C+} when He$^+$ dissociates \ce{CO}. He$^+$ is produced either by X-rays or cosmic rays depending on which dominates in the surface layers. There are two pathways forming \ce{C2H2} as shown in the two color-shaded regions of Fig.~\ref{UMIST+235}: the neutral-neutral pathway (purple) and the ion-molecule pathway (orange). The formation of \ce{C2H2} via the neutral-neutral pathway proceeds via the hydrogenation of C. The reaction of \ce{H2} with C forms \ce{CH2} which dissociates to form CH. The reaction of C with H also forms CH. 
These reactions are barrier less and so depend only on the abundances of H, \ce{H2} and C. Once CH is formed, it can form \ce{C2} via addition of neutral C. This is followed by abstraction of H to form C$_2$H and finally C$_2$H$_2$. 
The ion-molecule pathway proceeds through hydrogen addition. Starting from C$^+$, this leads to the formation of \ce{CH3+} via \ce{CH+}. The addition of neutral C then forms \ce{C2H+}. Again, as these reactions are barrier less, the availability of neutral C is the limiting factor here. Thus for both pathways, the availability of C, H and \ce{H2} is pivotal in deciding the formation route of \ce{C2H2} in the surface layers. This fundamental chemistry is valid for all the models. The pathways leading to \ce{C2H2} via C and \ce{C+} are same in all the models. The rates of the reactions differ because of the differences in the abundances and the rate coefficients between the two databases.

\ce{C2H2} is destroyed to form higher hydrocarbons via the addition of C in the fiducial model. An example is the formation of \ce{C3H}. The other destruction path is \ce{C2H2} photodissociation to \ce{C2H}.

In the fiducial model, at grid point 2, we find the ion-molecule chemistry to dominate. The most important pathways forming C$_2$H$_2$ are:
\begin{equation} \label{a2}
    \ce{C3H2} + \ce{C2H3+} \rightarrow \ce{C2H2} + \ce{C3H3+}
\end{equation}
\begin{equation}
    \ce{C3H3+} + \ce{e-} \rightarrow \ce{CH} + \ce{C2H2}
\end{equation}
\begin{equation}
    \ce{C3H} + \ce{C2H3+} \rightarrow \ce{C2H2} + \ce{C3H2+} .
\end{equation}
These formation pathways are not relevant in the surface layers. As the surface layers are UV dominated, C$_3$H$_2$ is photodissociated to C$_3$H instead of forming C$_2$H$_2$. Figure~\ref{5_plots} shows the difference in the abundances for \ce{C2H2} in all our models which we discuss in the following subsections.

\subsection{The extended hydrocarbon chemistry using the UMIST database} \label{The extended hydrocarbon chemistry using the UMIST database}

In model 3, we add the hydrocarbon species that are common to the UMIST2012 and the KIDA2014 rate database as shown in Table~\ref{Table3}. Model 3 only uses reactions from the UMIST2012 rate database. The new \ce{C2H2} abundances are shown in Fig.~\ref{5_plots} (lower left panel). To understand the differences in abundance of C$_2$H$_2$ between the fiducial model and this model 3, Fig.~\ref{UMIST+317} highlights the differences in the pathways described in the previous section. 

In the surface layers, i.e.\ grid point 1, the total formation rate of C$_2$H$_2$ increases by $\sim$ 41\% in model 3 relative to the fiducial model. This increase is due to the new pathways that are active now and which were missing from the fiducial model. 
Figure~\ref{UMIST+317} shows these new pathways that form \ce{C2H2}.

The other difference is the formation of C$_2$H via breaking down larger hydrocarbons in model 3. For example, the following two reactions were absent in the fiducial model
\begin{equation}
    \ce{HC4H} + h\nu \rightarrow \ce{C4H} + \ce{H}
\end{equation}
\begin{equation}
    \ce{C4H} + h\nu \rightarrow \ce{C2H} + \ce{C2}.
\end{equation}
C$_4$H dissociates to smaller hydrocarbon molecules like \ce{C2H} and C$_2$. These two molecules are steps in the formation of C$_2$H$_2$ as seen in Fig~\ref{UMIST+235}. Enhancing their abundances will propagate and lead to a higher abundance of \ce{C2H2}.

The abundance of \ce{C2H2} in the outer midplane i.e.\ grid point 2 drops by 6 orders of magnitude in model 3 with respect to the fiducial model (Fig.~\ref{5_plots}). Due to the presence of longer hydrocarbons and their corresponding more stable ices, the carbon is bound in the ices of longer hydrocarbons. The most abundant ice at this location is benzene ice C$_6$H$_6$\#. The gas-phase longer hydrocarbon species are concentrated in the inner midplane regions roughly inside 0.5~au at $T_{\rm gas}$ higher than $\sim$200~K. 

In the inner midplane, i.e.\ grid point 3, the total formation rate of C$_2$H$_2$ decreases by an order of magnitude in model 3 compared to the fiducial model resulting in a decrease in abundance by an order of magnitude. The dominant formation pathway in the fiducial model is 
\begin{equation}\label{1_h}
    \ce{C2H3} + \ce{H} \rightarrow \ce{C2H2} + \ce{H2}.
\end{equation}
This reaction is exactly balanced by the destruction of \ce{C2H2} via a three body reaction 
\begin{equation}\label{3b1}
    \ce{C2H2} + \ce{H} + \ce{M} \rightarrow \ce{C2H3} + \ce{M}
\end{equation}
in model 3 making the thermal decomposition of \ce{CH2CCH} by M (where M can be H, \ce{He} or \ce{H2}) the dominant formation pathway. As we go deeper in the disk ($r\,=\,0.07\rm au$, $z\,=\,0.007\rm au$, $T_{\rm gas\,}$=\,1100\,K and n$_{\rm <H>}\,$=\,$2.3\cdot10^{14}\rm cm^{-3}$), we find that these major formation pathways are balanced by neutral-neutral or three body destruction pathways making secondary species like \ce{H2O} important in forming \ce{C2H2}.

The neutral-neutral destruction reaction of \ce{C2H2}
\begin{equation}\label{a1}
    \ce{C2H2} + \ce{CH2} \rightarrow \ce{CH2CCH} + \ce{H}
\end{equation}
is favoured over cosmic ray induced photodissociation of \ce{C2H2} to form \ce{C2H}; this reaction is missing from the fiducial model. This again highlights the importance of adding higher hydrocarbons to determine more reliable \ce{C2H2} abundances.

\subsection{Role of isomers in the extended hydrocarbon network}

We compare models\,2 and 4 that is the large DIANA and extended chemical network using the KIDA2014 rate database, respectively. We find two new pathways which are reaction\,\ref{3} and the dissociative recombination of \ce{CH2CCH+} with \ce{e-} forming \ce{C2H2} in model 4 (at point 1, Fig\,\ref{KIDA+317}). These are the new species added in the extended network. 
The dissociative recombination reaction of \ce{C3H2} with \ce{e-} becomes less important in model 4.
This is due to a lowered abundance of \ce{C3H2+}. Its formation proceeds through
\begin{equation}
   \ce{H2} + \ce{C3H+} \rightarrow \ce{C3H3+} + h\nu
\end{equation}
\begin{equation}
    \ce{H} + \ce{C3H3+} \rightarrow \ce{C3H2+} + \ce{H2}
\end{equation} in model 2.
In the extended network, the first step can branch to two isomers, \ce{C3H3+} (\ce{c-C3H3+}) and \ce{CH2CCH+} (\ce{l-C3H3+}) thus lowering the abundance of \ce{C3H3+} consequently lowering the abundance of \ce{C3H2+}. However, the decrease in this rate is compensated by the two new pathways above, overall resulting in an increase of $\sim$ 40\% in the abundance of \ce{C2H2} at grid point 1 in model 4.

One of the major formation pathways in both models at grid point 3 is
\begin{equation}
    \ce{C2H3+} + \ce{H2O} \rightarrow \ce{C2H2} + \ce{H3O+} .
\end{equation} 
The thermal decomposition of \ce{CH2CCH} by M and dissociative recombination of \ce{C5H5+} with \ce{e-} add to the production of \ce{C2H2}. Both were missing from the DIANA chemical network thus increasing the \ce{C2H2} abundance by 2 orders of magnitude in the inner disk midplane region (point 3).

\begin{figure*}
    \centering
    \includegraphics[width=\linewidth]{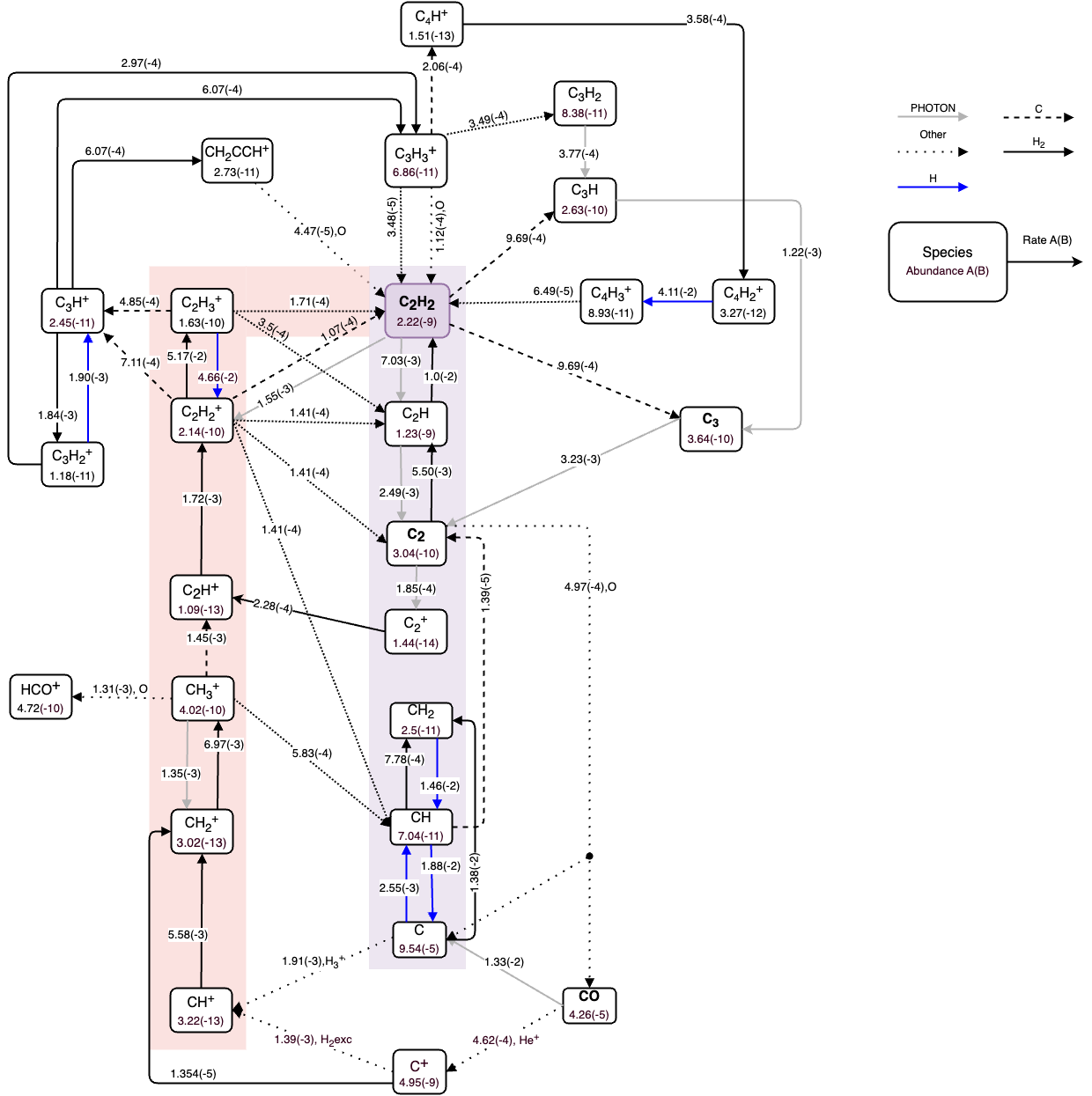}%
    \caption{The chemical network for the surface layers for model~5 (final hydrocarbon network) centered around \ce{C2H2}.}
    \label{327_network}
\end{figure*}

\subsection{Comparison between the UMIST and KIDA rate databases} \label{Comparison between the UMIST and KIDA rate databases}
Figure~\ref{5_plots} shows the difference in the radial extent of the surface layer of \ce{C2H2} in models using the UMIST2012 and the KIDA2014 rate database. We traced this back to the layer with high \ce{H2} abundances extending down to smaller radii in models using the KIDA2014 database. The chemical pathways displayed in Fig.~\ref{UMIST+235} show that \ce{H2} is crucial in the formation of \ce{C2H2}. 
Comparing two models using the same chemical network but different rate databases (models 3 and 4), the vertical height at which H/H$_2$ transition occurs is higher by $\sim$0.005\,au at the radial distance of 1~au in model 4 (KIDA2014) compared to model 3 (UMIST2012). 
Models using the UMIST2012 database (fiducial and model 2) have a higher destruction rate of \ce{H2}. There is more destruction of H$_2$ by atomic oxygen in the UMIST2012 model compared to the KIDA2014 model due to different rate coefficients for the reaction
\begin{equation}
    \ce{O} + \ce{H2} \rightarrow \ce{H} + \ce{OH}.
\end{equation}

The \ce{C2H2} abundance at grid point 3 for model 2 is lower compared to the fiducial model as seen in Fig~\ref{5_plots}. In the fiducial model, C is locked in \ce{C2H2} while in model 2 more of the C is in instead \ce{CH4} at the inner midplane (grid point 3). We attribute this difference to missing reactions and different rate coefficients given in the databases. The reaction
\begin{equation}
    \ce{CH_3} + \ce{C_2H_3} \rightarrow \ce{C_2H_2} + \ce{CH_4}
\end{equation}
is missing from the KIDA2014 database, but it is one of the dominant destruction reactions for \ce{C2H3} and the dominant formation reaction for CH$_4$ and \ce{C2H2} when using UMIST2012 (fiducial model). The following dominant formation reaction in the fiducial model 
\begin{equation}
    \ce{C2H} +\ce{H2} \rightarrow \ce{C2H2} + \ce{H}
\end{equation} 
has a barrier of 1300\,K in KIDA2014 but only 130\,K in UMIST2012 explaining the lower rate in model 2.
Another important species contributing to the formation of \ce{C2H2} is \ce{C2H3+}. This molecular ion is formed via the reaction
\begin{equation}
\ce{C2H2+} + \ce{H2} \rightarrow \ce{C2H3+} + \ce{H} 
\end{equation}
which is a barrierless reaction in UMIST2012 but has an activation energy of 2000\,K in KIDA2014. The combination of these differences lead to a lower abundance of \ce{C2H2} in model 2 at grid point 3 when compared to the fiducial model.

\subsection{Benzene formation}

Benzene is confined to the inner midplane below $ r\,\leq 0.2$~au and a $z/r$ of $\sim\,0.1$. It also appears in a narrower radial region in model~4 (KIDA2014, extended hydrocarbon chemistry). To find the dominant formation and destruction pathways in models, we analyse the chemistry in the inner midplane at $ r\, =\, 0.086$\,au and $ z/r \,=0.07$ with $T_{\rm gas}$\,=\,$T_{\rm dust}$\,=\,570\,K in an optically thick region with $A_V^{\rm ver}$\,=\,150 and $A_V^{\rm rad}$\,=\,2400 and $n_{\rm <H>}$= $5.6\cdot10^{14}\rm cm^{-3}$. We choose this point as it has the maximum abundance of benzene in model 5 (final extended chemical network). 
The network might be missing some reaction as the databases are not complete in hydrocarbons with 6 C atoms such as C$_6$H$_4$, C$_6$H$_5$ etc. 
The pathway to form benzene is via \ce{CH4} and \ce{C2H2}. In model 3 (using UMIST2012), the dominant pathway of formation is 
\begin{equation}\label{b3}
    \ce{CH4} + \ce{C5H2+} \rightarrow \ce{C6H5+} + \ce{H}
\end{equation} contributing $\sim$90\% to the total formation rate followed by
\begin{equation}\label{b4}
    \ce{C2H2} + \ce{C4H3+} \rightarrow \ce{C6H5+} + h\nu
\end{equation} contributing only $\sim$5\%. \ce{C6H5+} reacts further with \ce{H2} to form \ce{C6H7+} which subsequently recombines with an electron to form benzene
\begin{equation}\label{b1}
    \ce{H2} + \ce{C6H5+} \rightarrow \ce{C6H7+} + h\nu
\end{equation}
\begin{equation}\label{main}
    \ce{C6H7+} + \ce{e-} \rightarrow \ce{C6H6} + \ce{H} .
\end{equation}
Model 4 (KIDA2014) also follows the same formation scheme as explained above. However, the abundance of benzene at this location in model 4 is of the order of 10$^{-14}$ but even when analysing the region with the highest abundance (10$^{-10}$) we find the same formation mechanism which then leads to \ce{C6H6} via dissociative recombination (reaction\,\ref{main}).

\subsection{The final hydrocarbon network}

Based on what we learned from the previous comparisons, we compiled the final chemical network that has 327 species in total combining the large DIANA network with 235 species and the 92 additional hydrocarbons from Table~\ref{3}. The leading database is UMIST2012 as it has more species than KIDA2014 and we add a few reactions picked from the KIDA2014 database as described in Sect.~\ref{Building the Chemical network}. The network includes the three body and thermal decomposition reactions as mentioned in Appendix\,\ref{3body reactions}.

The basic formation and destruction pathways of \ce{C2H2} remain the same as described in Sect.\,\ref{The extended hydrocarbon chemistry using the UMIST database}. Figure~\ref{327_network} details the chemical pathways in the surface layers of the disk. The dominant reactions forming \ce{C2H2} are reactions~\ref{1},
\begin{equation} \label{4}
    \ce{C2H} + \ce{H2} \rightarrow \ce{C2H2} + \ce{H} \,\,\, and
\end{equation}
\begin{equation}\label{20}
    \ce{C2H3+} + \ce{e-} \rightarrow \ce{C2H2} + \ce{H}.
\end{equation}
These were also the dominant formation reactions in model 3 for the surface layers. The addition of new hydrocarbons changes the abundances of \ce{C2H2} in the inner and outer midplane by 2 and 7 orders of magnitude, respectively, compared to the fiducial model for the reasons explained in previous sections. But it does not affect the \ce{C2H2} abundances in the surface layers much (increase by a factor of $\sim$1.5).

In the final network, the reaction of \ce{H2} with \ce{C6H5+} leads to \ce{C6H7+} that recombines with an electron to form benzene (see reactions \ref{main} and \ref{b1}). Reaction~\ref{b3} contributes $\sim$83\% and reaction~\ref{b4} contributes $\sim$16\% to the total formation rate of \ce{C6H5+}. Hence, both \ce{CH4} and \ce{C2H2} contribute to the formation of \ce{C6H6} with the route through \ce{CH4} being the dominant one. Figures \ref{HC_M} and \ref{HC_M1} show the effect of adding complex hydrocarbons and highlights how carbon chain length changes throughout the disk. The shades of blue and red corresponds to the single C and two C atom bearing species. The diversity in the ices also increase on using the extended chemical network. This network will be used for the remainder of the paper.

\section{Mid-IR spectra}\label{Mid-IR spectra}

\begin{figure}
    \begin{flushleft}
    \includegraphics[width=\linewidth]{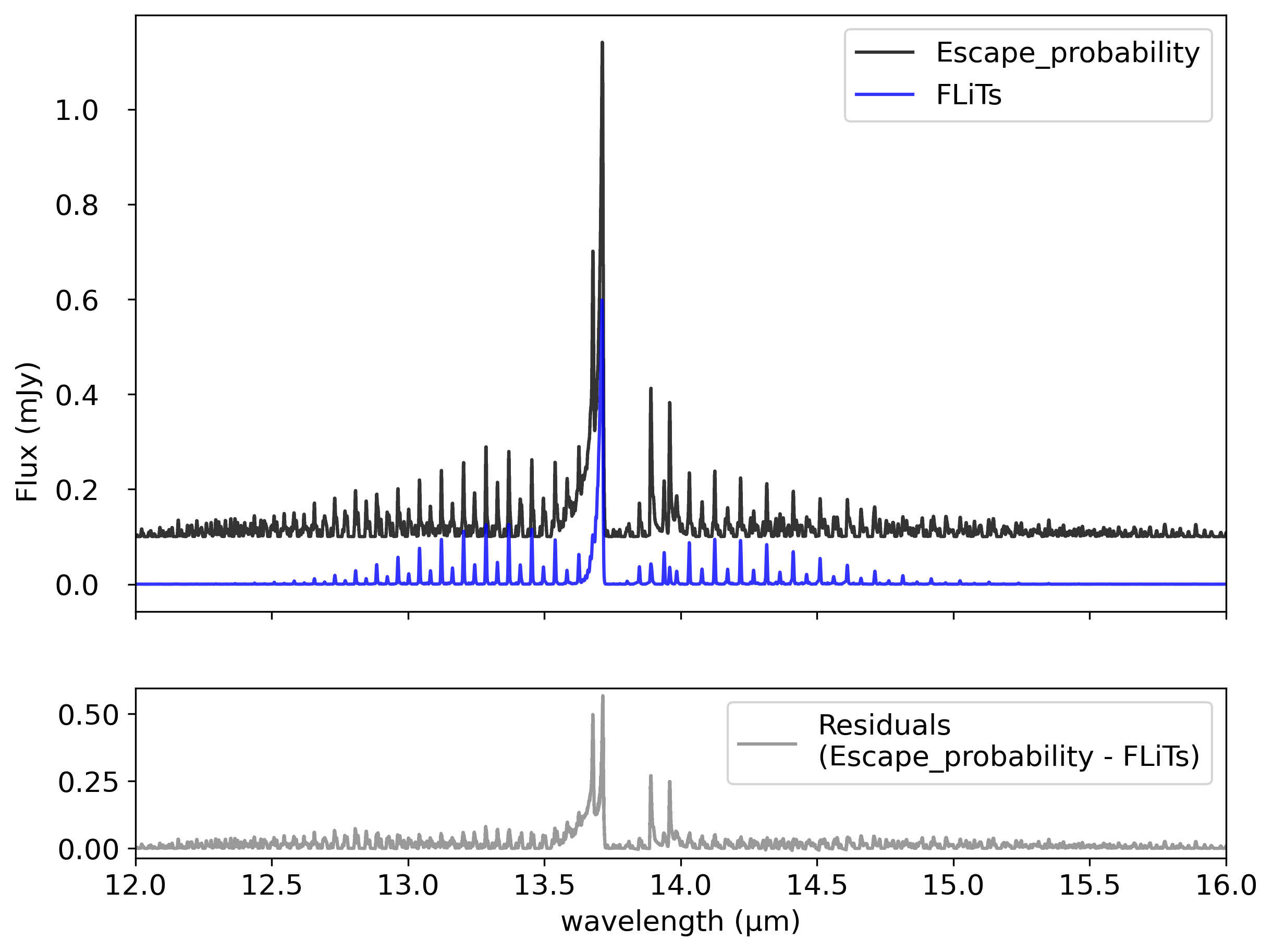}
    \caption{The flux emitted by \ce{C2H2} in model 5, convolved at a spectral resolution of $R\,=\,3000$ calculated using the escape probability method (black) and using the FliTs code (blue). The bottom panel shows the differences in the line flux between the two methods.}
    \label{flits_compare}
    \end{flushleft}
\end{figure}

\begin{figure}
    \begin{flushleft} 
    \includegraphics[width=\linewidth]{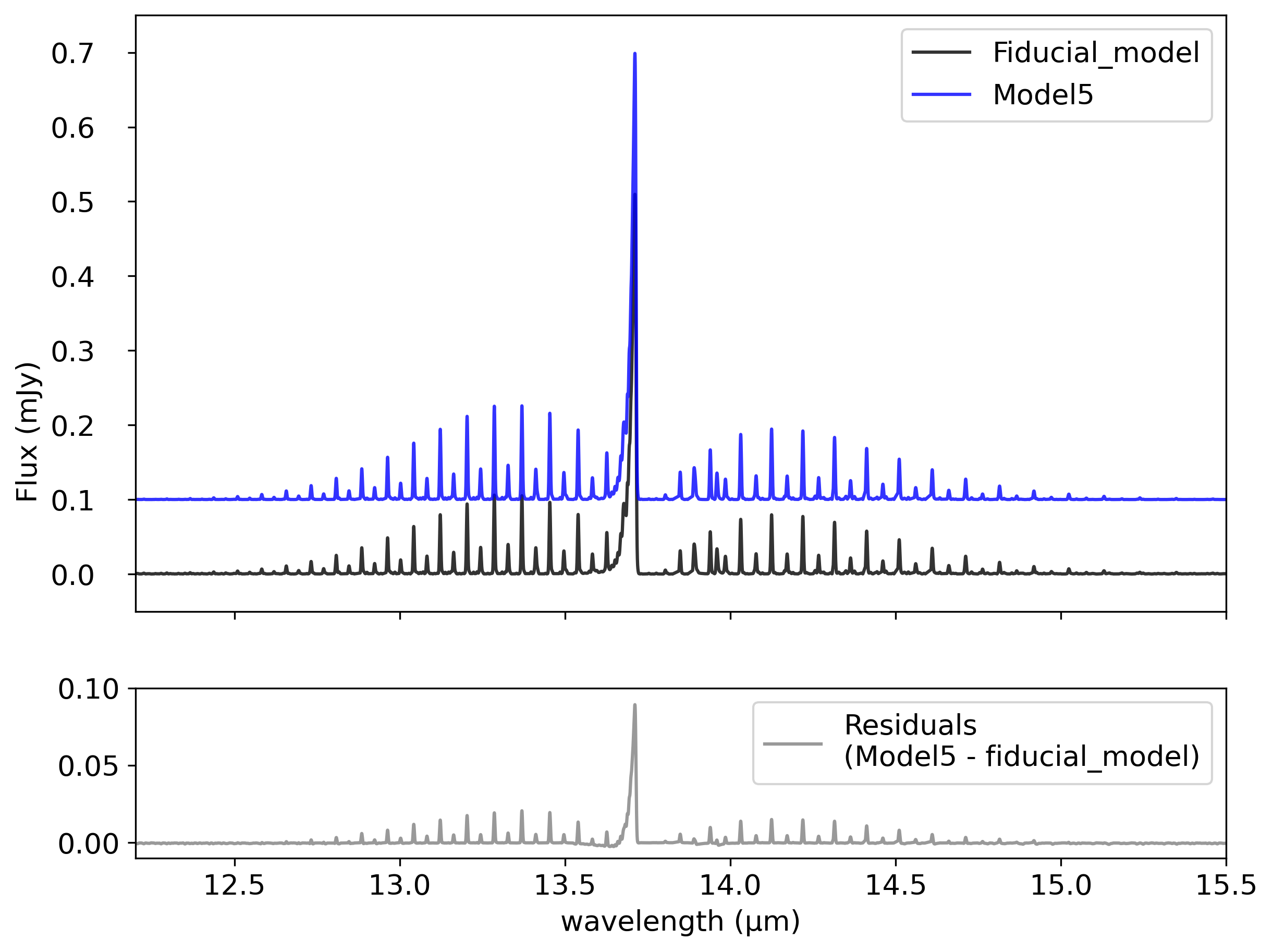}%
    \caption{\ce{C2H2} flux using the FLiTs code for the model with large DIANA network (fiducial model) and the extended hydrocarbon chemistry (model~5) convolved to the spectral resolution of $R\,=\,3000.$ The bottom panel shows the difference in the line flux between the two models.}
    \label{flits}
    \end{flushleft}
\end{figure}

\begin{figure}
    \centering
    \includegraphics[width=\linewidth]{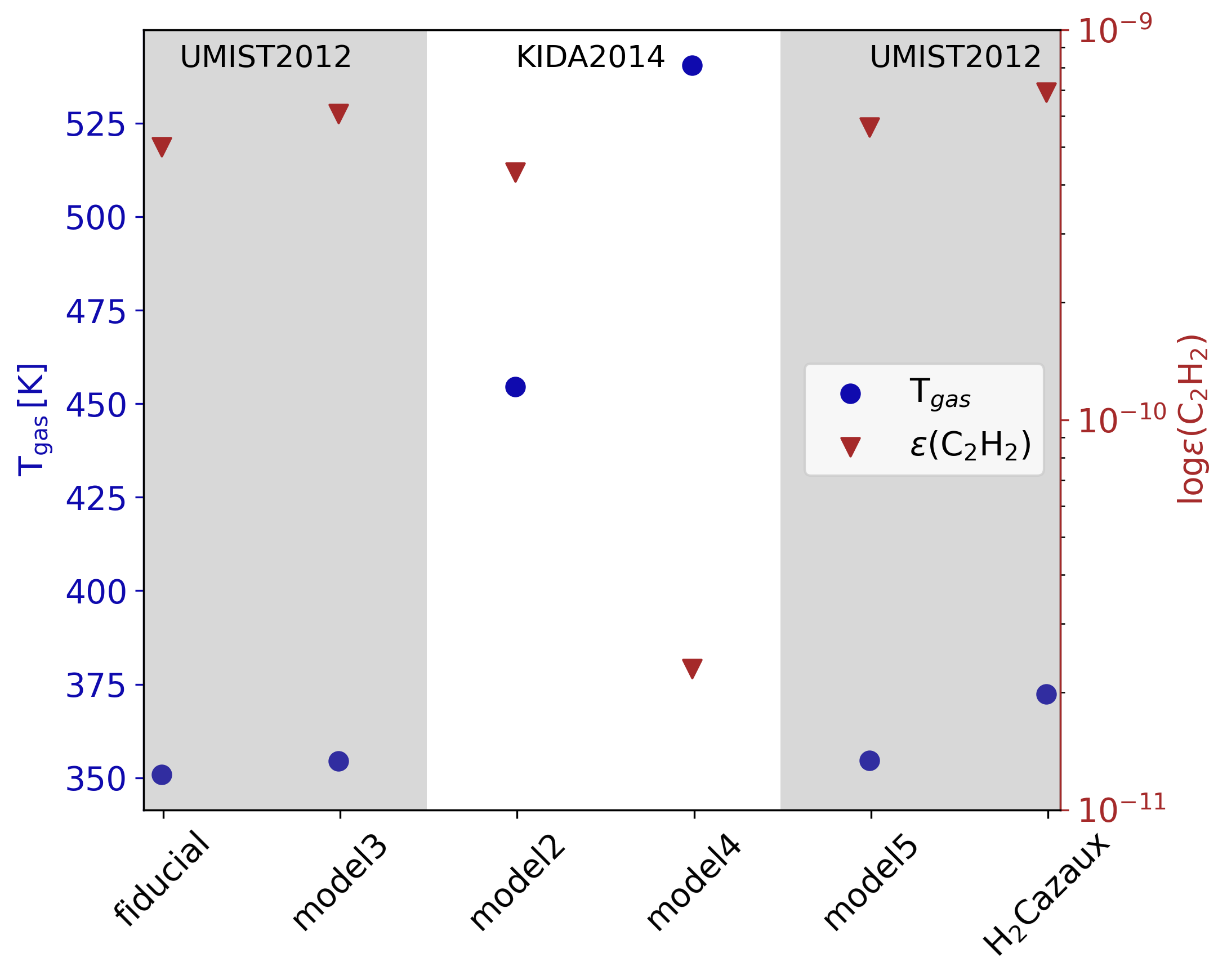}
    \caption{The gas mass weighted abundance of \ce{C2H2} (brown) and \ce{C2H2} abundance weighted gas temperature (blue) of the emitting region of \ce{C2H2} at 13.71\,$\mu$m for all the models. The gray background depicts the models using the UMIST2012 database and white the ones using the KIDA2014 database.}
    \label{gas_temp}
\end{figure}

\begin{figure}
    \centering
    \includegraphics[width=\linewidth]{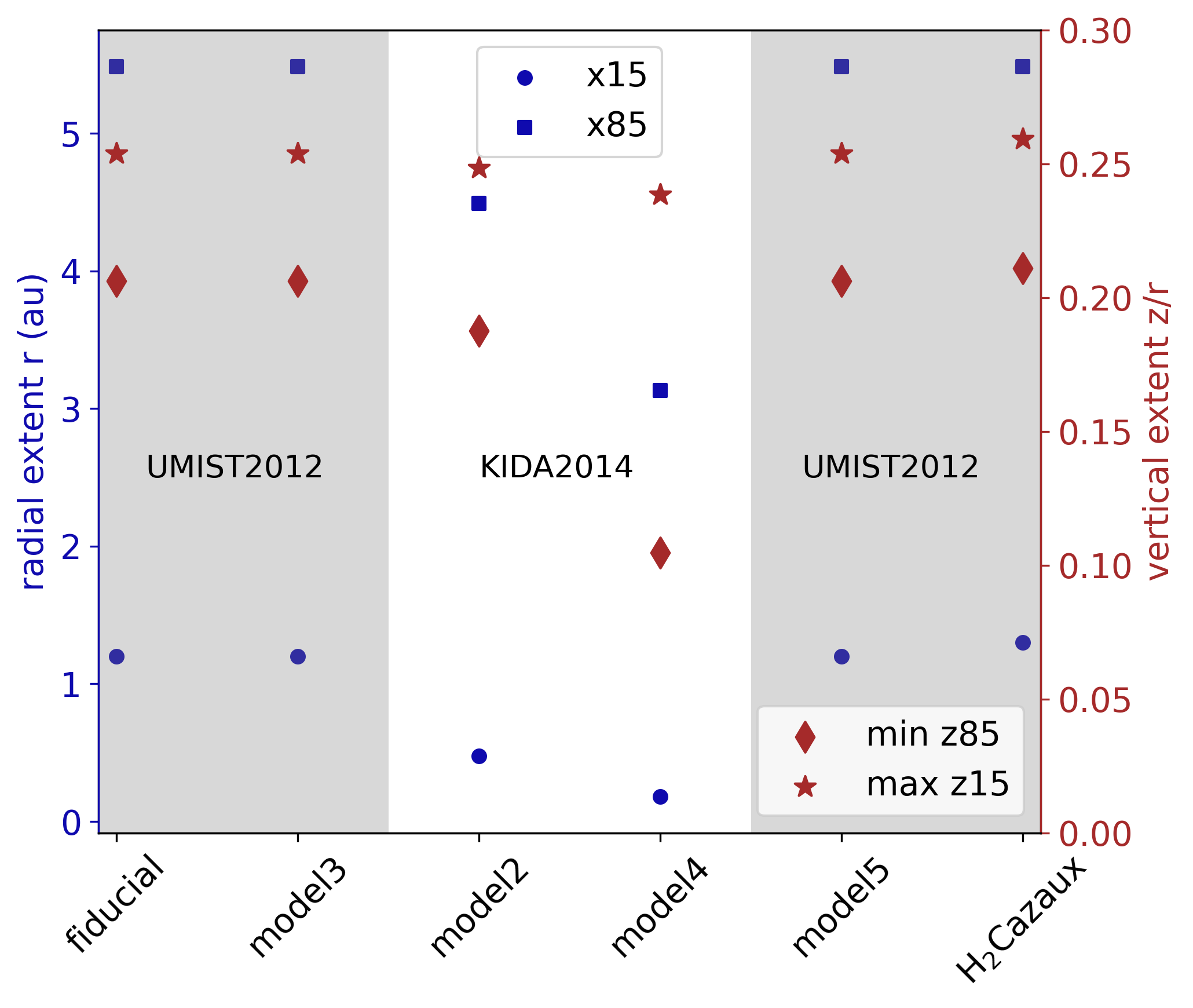}
    \caption{The effect on the emitting region of the \ce{C2H2} at 13.71\,$\mu$m line due to different chemical networks and the rate databases. The left and right axes depict the maximum radial and vertical extent of the emitting region. x15, x85, z15(r) and z85(r) denote the radii and heights at which 15 and 85\% of the line flux originate, respectively. Min z85 and max z15 indicate the maximum vertical extent (z/r) of the region. The gray background is used to highlight the models using the UMIST2012 database, while the white background pertains to models using the KIDA2014 database.}
    \label{emitting_region}
\end{figure}

\begin{figure}
    \centering
    \includegraphics[width=\linewidth]{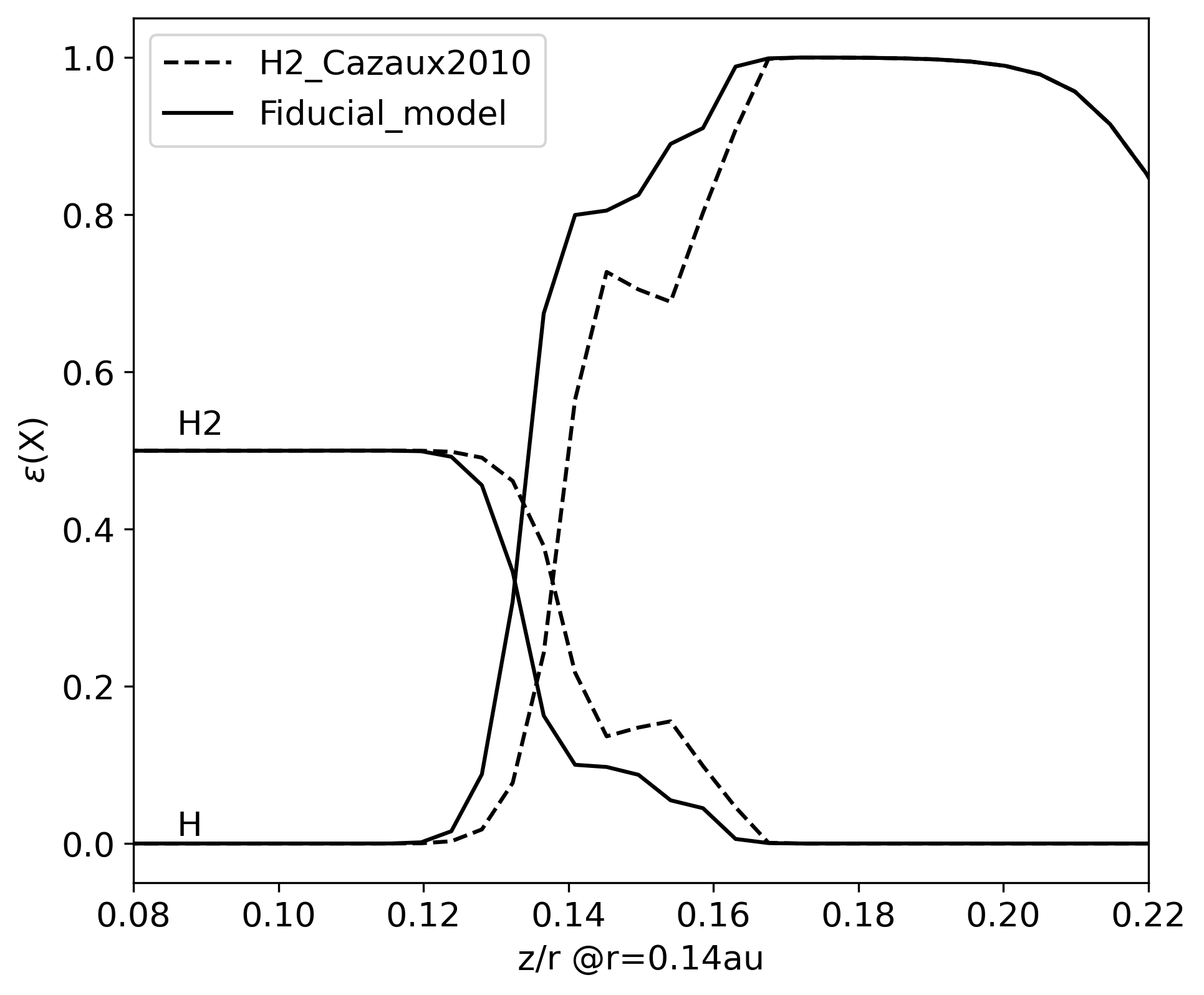}
    \caption{The effect of using updated H$_2$ formation rate of \cite{Cazaux2010} compared to \citeyear{Cazaux2004} on the H$\rightarrow$ H$_2$ transition layer. The figure compares the fiducial model using the large DIANA chemistry with the disk having the extended chemical network using the formation mechanism and the energy barrier reported in \cite{Cazaux2010}.}
    \label{H2H}
\end{figure}

The extended hydrocarbon chemistry has an effect on the chemical abundances, especially on \ce{C2H2}. We study here whether this has an impact on the mid-infrared spectra emitted from the disk. We use the escape probability method which yields vertically emitted total line fluxes \citep{Woitke2009} and FLiTs developed by Michiel Min \citep{Woitke2018}. FLiTs does a full line radiative transfer including the dust continuum and line opacity overlap. We use HITRAN 2009 \citep{Rothman2009} as our spectroscopic database. The rules for selecting the line transitions are taken from \cite{Woitke2018}.

Figure~\ref{flits_compare} shows the line flux for \ce{C2H2} calculated using the escape probability method and FLiTs convolved to a resolution of 3000 appropriate for JWST. The weak lines at $\sim14\,\rm \mu$m have higher fluxes using the escape probability method compared to FLiTs (see residuals in Fig.\,\ref{flits_compare}). The escape probability treatment do not take into account absorption by the continuum which can be an issue for molecules emitting from relatively low heights in the disk. The escape probability produces a factor of 2 higher integrated \ce{C2H2} (11.8-16.22\,$\rm \mu$m) flux relative to FLiTs as shown in Fig~\ref{flits_compare}. In the following, we use FLiTs to calculate the spectra.  

\subsection{Implication on mid-IR spectra}\label{4.1sect}
Figure~\ref{flits} shows the flux emitted by \ce{C2H2} for the model with the expanded hydrocarbon chemistry (model~5) and the one using the large DIANA chemical network (fiducial model). The total integrated flux in model 5 is higher by 10\%  relative to the fiducial model. This increase is solely because of the use of the expanded chemical network as we fixed the gas temperature of the disk. The peak fluxes produced in model 5 are $\sim0.6$\,mJy at $R=3000$ and $\sim0.33$\,mJy at $R=600$. The peak flux for \ce{C2H2} in T Tauri disks as observed by Spitzer is $\sim$20\,mJy at a $R=600$. We find that using an extended chemical network cannot enhance \ce{C2H2} emission to the typical flux level of mid-infrared observations of TTauri disks. A wide exploration of disk physical parameters like UV radiation, disk mass etc. using this network is now required to identify the key parameters and processes that allow us to reproduce the observed fluxes.

The integrated line flux for \ce{C2H2} is a factor 3 higher in model 4 (KIDA2014) compared to model 3 (UMIST2012) that have the same chemical network (i.e. equal number of species). The difference arises from the different rate coefficients in the two databases. This leads to the molecular layer being shifted radially inwards in model 4 and being $\sim$~50\% more vertically extended (see Fig.\,\ref{emitting_region}, and also Sect.~\ref{Comparison between the UMIST and KIDA rate databases}). As a result, the \ce{C2H2} abundance and the abundance weighted gas temperature is higher in model 4 by $\sim$~53\% (see Fig.\,\ref{gas_temp}), which leads to the factor 3 higher line fluxes.

As seen in Fig.\,\ref{fig_63}, in our fiducial model (TTauri disk model, power law surface density profile with a sharp inner edge), only the surface layers contribute towards the mid-IR emission. However, Woitke et al. (submitted) propose a disk model for EX Lupi with a steadily increasing surface density at the inner rim. This fits the spectral energy distribution, the overall shape of the mid-IR spectra observed by JWST, and the observed molecular features and its characteristics like emitting area and gas temperatures. In their model, they find a high abundance of \ce{C2H2} in the O-rich gas around the inner rim, caused by X-ray irradiation. This region corresponds to our inner midplane (grid point 3) reservoir of \ce{C2H2}. This recent work shows that for models with a different physical structure, the inner regions (grid point 3) can also contribute to the mid-IR emission.

\subsection{The effect of \ce{H2} formation on \ce{C2H2}} \label{Effect of}

In Sect.~\ref{investigating the hydrocarbon chemistry} we show that the H/H$_2$ transition is important for the abundance of \ce{C2H2} and in Sect.\,\ref{4.1sect} we show that this affects also the flux emitted from this molecule. Hence, the formalism of how H$_2$ forms may affect the abundance of \ce{C2H2} in the disk surface layer. 

To investigate this, we compare the two approaches of H$_2$ formation described in \cite{Cazaux2004} and \cite{Cazaux2010}. In cold regions, the formation of H$_2$ from physisorbed H dominates whereas in warm regions, the formation of H$_2$ can also proceed via chemisorbed H. Figure~\ref{H2H} shows how the height of the H/H$_2$ transition ($z/r$) shifts upwards by 0.005~au at r\,=\,1\,au when using an improved treatment of mobility of H atoms on grain surfaces at warmer temperatures. Model \ce{H2}Cazaux in Figs.\,\ref{gas_temp} and \ref{emitting_region} show the effect on the gas temperature, \ce{C2H2} abundance and the emitting area of using different databases and the chemical networks. Model \ce{H2}Cazaux used extended chemical network with the UMIST2012 rate database. The peak flux emitted by \ce{C2H2} at $\sim13.7\,\rm \mu$m (Q branch) increases by a factor of $\sim$2 (escape probability method) when using the \cite{Cazaux2010} formalism. The increase in flux can be attributed to the increased gas temperature of the line emitting region as shown in Fig.\,\ref{gas_temp} when compared to model 5. Also in this case, the surface layer of \ce{C2H2} extends radially inwards into the inner disk regions as warm temperatures promote the chemisorption channel for \ce{H2} formation.

\section{Discussion}\label{Discussion}
We studied various pathways leading to the larger hydrocarbons in protoplanetary disks, specifically \ce{C2H2}. We isolate the major formation and destruction pathways for \ce{C2H2} and \ce{C6H6} and their differences on the basis of whether the UMIST2012 or KIDA2014 rate databases are used. We study the effect of the H/\ce{H2} transition layer on the surface layers of \ce{C2H2} when using different \ce{H2} formation formalism. We compare in the following the hydrocarbon chemistry from our network in disks to earlier works on hot cores, disks, AGB stars and planetary atmospheres. 
\subsection{Acetylene}
Comparing to previous disk studies, we find that the mechanism to unlock C from CO is the same as described in \citet[and references therein]{Bast2013} and \cite{Walsh2015}: CO is either photodissosciated to C or reacts with \ce{He+} produced by CR or X-ray photons (depending on which one dominates) to form \ce{C+}. The chemistry pathways to form \ce{C2H2} shown in these studies are similar to what we find. \cite{Walsh2015} do not find the neutral-neutral formation pathway to \ce{C2H} from \ce{C2} important. 
Our chemistry is also similar to that in \cite{Agundez2008}, although their models begin to form CH by C abstracting H from \ce{H2}. This reaction has a high energy barrier (12000\,K) and is also reported in \cite{Walsh2015}. Despite having this reaction in our disk models, we form CH via radiative association of C and H and destruction of \ce{CH2} which is a barrier-free reaction and has no temperature dependence. This difference arises because \cite{Agundez2008} follow the time dependent evolution of chemistry with hydrogen being initially molecular.

Warm carbon chain chemistry (WCCC) sources are Class 0/I objects that contain hydrocarbons (e.g. L1527) \citep{Sakai2008}. 
Densities in WCCCs are of the order of $10^{9} \rm cm^{-3}$ and temperatures are $\sim\,300$~K. The chemistry is initiated by sublimating methane ice, which subsequently reacts with \ce{C+} to form hydrocarbon molecules \citep{Sakai2013}. The formation of higher hydrocarbons proceeds via addition of \ce{C+} followed by dissosciative recombination with \ce{e-} or via reaction with \ce{CH4} \citep{Sakai2013}. The chemistry is completely dominated by ion-neutral chemistry. In our disk models (point 1 which has similar density, temperatures as WCCC sources (see Fig.\,{\ref{struc.pdf}})), we find reactions with C to dominate over reactions with \ce{C+}. Higher hydrocarbons are formed also via addition of neutral \ce{C}. Addition of \ce{H2} to ions is the main pathway in disks to form higher hydrogenated ion contrary to the addition of \ce{CH4} to molecular ions which occurs in WCCC sources. The high abundances of \ce{C+} required in WCCC sources are maintained either by high cosmic-ray induced ionization ($\zeta$) or high UV. WCCCs can have $\zeta \, \geq \,\rm10^{-16} \rm s^{-1}$ \citep{Kalvans2021} whereas we use $\zeta \,= \,\rm10^{-17} \rm s^{-1}$ in our disk models. We also use a steady state chemistry model while the WCCC models study time dependent chemistry starting from cold icy conditions in dark cores and warming up over short timescales as gravitational collapse provides energy. So there is also a fundamental difference in timescales of chemistry and history of material.

C-rich Asymptotic giant branch (AGB) stars have been studied in detail by for example \cite{Cherchneff1993,Millar1994}. Typical gas temperatures in the inner circumstellar envelopes around AGB stars are $\sim$1000\,K and densities are the order of 10$^{10} \rm cm^{-3}$. The neutral-neutral chemistry dominates in these objects while the chemistry in disk surface layers that have similar physical conditions (see Fig.\,\ref{struc.pdf}) is a mixture of ion-neutral and neutral-neutral chemistry. The difference in the initial abundances between our disk model and that of \citet[][Table 2]{1993ApJ...410..188C} prevents a direct comparison. The trend of odd-even effect where even numbered C hydrocarbons are more abundant than the odd C \cite[see][Fig.~3]{Cherchneff1993} is also seen in our disk models.

\cite{Loison2019,Vuitton2019} studied the hydrocarbon chemistry in atmospheres of planets and moons. The chemical networks in these studies take into account the termolecular and pressure dependent reactions. The temperatures range from $\sim$ 80 to 180\,K and pressures vary from 10$^{5}$ (surface) to 10$^{-7}$ Pascal with altitude \citep{horst2017}. These pressures correspond to a range of $\sim$ 10$^8$ to 10$^{20}\,\rm cm^{-3}$ of particle densities. The reactions destroying \ce{C2H2} to form \ce{C2H} and \ce{C2} in atmospheres \citep{Vuitton2019} are also dominant in the disk surface layers but the physical conditions are not the same, thus preventing an in depth comparison. 
\subsection{Benzene}
The reactions \ref{b4}, \ref{b1} and \ref{main} forming \ce{C6H6} are the dominant pathways in both studies \citep{Loison2019, Vuitton2019}. We find the same reactions to be the major pathways forming benzene in the disk. The reaction 
\begin{equation} 
    \ce{C3H3} + \ce{C3H3} + M \rightarrow \ce{C6H6} + M
\end{equation}
is found to contribute to form benzene at pressures $\sim$ 10$^{15}$ to 10$^{19}\,\rm cm^{-3}$ and at temperatures of $\sim$ 160\,K in atmospheres \citep{Loison2019}. However, \cite{Woods2007} who include this reaction, did not find it to be important in disks and it is therefore not included in our network. 

On comparing the benzene formation routes in models using the UMIST2012 and KIDA2014 database the formation scheme via reactions \ref{b4} and \ref{b1} leading to \ce{C6H6} are reported in \cite{McEwan1999} for the interstellar medium and are dominant in both the models. The reaction scheme forming benzene in protostellar environments outlined in \cite{Woods2007} as
\begin{equation}\label{abc}
    \ce{CH3CCH} + \ce{C3H4+} \rightarrow \ce{C6H7+} + \ce{H}
\end{equation}
is not dominant in any of our models. The thermal decomposition reaction of \ce{CH2CCH} with M to form \ce{C2H2} becomes dominant and is favoured over the neutral-neutral reaction of \ce{CH2CCH} with \ce{H2} forming \ce{CH3CCH}. This decreases the abundance of \ce{CH3CCH} resulting in reaction \ref{abc} not being dominant.

\subsection{Future work}
A chemical network can never be complete. Based on the species for which the reactions of formation and destruction were available, the extended network for hydrocarbon chemistry is provided. The difference in rate coefficients from the rate databases affects the chemistry and the abundance of species in the disk, but we show that the impact on the observable \ce{C2H2} line fluxes is lower than a factor 2. 

The values of adsorption energies shift the ice lines and the uncertainties in these thus affect the chemistry. Laboratory measurements of enthalpy of formation for a number of higher hydrocarbon ions are needed as they can also affect the chemical structure of the disk. We also miss species from the chemical network as we are limited by the rate databases. Our network does not include \ce{C6H5} or \ce{C6H4} (as explained in Sect.\,\ref{Building the Chemical network}).

A clear distinction between the isotopomers of species and their rate coefficients needs to be included in the rate databases. KIDA has more species compared to KIDA2014, but it is not in the format of an easy to use ratefile. Often multiple rate coefficients exists and so the user has to decide which rates are better. In addition, the database changes constantly, causing an issue to reproduce results at a later stage. These are fundamental differences to UMIST which clearly recommends rates and has stable releases. Both approaches of course have their pros and cons. 

We lack currently the framework to fully take into account the pressure-dependent reactions in {P{\small RO}D{\small I}M{\small O}} which are dominant in planetary atmospheres. Adding these is beyond the scope of this paper and will be the subject of a future study. These reactions might be important in the midplane (point 3) but less so in the surface layers of the disks. It is necessary to test in the future the importance of such reactions for the high density inner disks as we may miss some key pathways.

\section{Conclusion}
In this paper, we aim to understand the hydrocarbon chemistry in the inner, warm and dense regions of planet-forming disks around T Tauri stars. We present an extended chemical network that includes hydrocarbon species with up to 8 carbon atoms. It includes the linear and cyclic isotopomers of several species, limited by the availability of such data in UMIST2012 and KIDA2014. With this network we can form the simplest cyclic hydrocarbon: Benzene. 

We find that a more extended chemistry alone does not explain the high fluxes observed by Spitzer. 
A more detailed study varying physical parameters of the disk using this chemical network is needed to better interpret the observations. We find that the abundances of \ce{C2H2} using the extended chemical network increase only by $\sim$40\% relative to the large DIANA chemical network in the surface layers (grid point 1). The reservoir of gas-phase \ce{C2H2} in the outer midplane disappears to form ices of longer hydrocarbons. The abundance in those regions hence drops by 7 orders of magnitude (grid point 2). There is a decreasing gradient of \ce{C2H2} abundance in the inner midplane which is more prominent in model using the extended chemical network (model\,5). With the expanded network, we find that in extremely inner regions of the disk the major \ce{C2H2} formation pathways were balanced by the neutral-neutral, three body or thermal decomposition destruction pathways making secondary species like \ce{H2O}, \ce{NH3} etc important. A detailed analysis of the three-body reactions network is needed and will be the focus of a subsequent paper.

H, \ce{H2}, C and \ce{C+} are crucial to form \ce{C2H2}. Therefore, the layer at which the H/\ce{H2} transition occurs is important and thus the details of the \ce{H2} formation mechanism. We find that this transition occurs higher up in the disk when using the updated formalism of \ce{H2} formation \citep{Cazaux2010}. This is also important in determining the radial extent of the surface layer where \ce{C2H2} is abundant.

The extended chemical network presented here, is important when studying longer hydrocarbons in the disks. With JWST discovering more complex hydrocarbon species in disks more complex chemical modelling is warranted. Our work presents thus a key step forward in modelling the hydrocarbons in disks for future comparison with observations from JWST.
\begin{acknowledgements}
This project has received funding from the European Union’s Horizon 2020 research and innovation programme under the Marie Sklodowska-Curie grant agreement No.
860470. CHR acknowledges the support of the Deutsche Forschungsgemeinschaft (DFG, German Research Foundation) research Unit "Transition discs" - 325594231. CHR is grateful for support from Max Planck Society. We thank the anonymous referee for their constructive comments.
\end{acknowledgements}

\bibliographystyle{aa}
\bibliography{papers}

\begin{thebibliography}{47}
\expandafter\ifx\csname natexlab\endcsname\relax\def\natexlab#1{#1}\fi

\bibitem[{A.~M.~Arabhavi(2023)}]{Arabhavi2023}
A.~M.~Arabhavi, I.~K. 2023, Science, submitted

\bibitem[{{Ag{\'u}ndez} {et~al.}(2008){Ag{\'u}ndez}, {Cernicharo}, \& {Goicoechea}}]{Agundez2008}
{Ag{\'u}ndez}, M., {Cernicharo}, J., \& {Goicoechea}, J.~R. 2008, \aap, 483, 831

\bibitem[{Anderson {et~al.}(2021)Anderson, Blake, Cleeves, Bergin, Zhang, Schwarz, Salyk, \& Bosman}]{Anderson2021}
Anderson, D.~E., Blake, G.~A., Cleeves, L.~I., {et~al.} 2021, The Astrophysical Journal, 909, 55

\bibitem[{{Bast} {et~al.}(2013){Bast}, {Lahuis}, {van Dishoeck}, \& {Tielens}}]{Bast2013}
{Bast}, J.~E., {Lahuis}, F., {van Dishoeck}, E.~F., \& {Tielens}, A.~G.~G.~M. 2013, \aap, 551, A118

\bibitem[{{Bergin} {et~al.}(2016){Bergin}, {Du}, {Cleeves}, {Blake}, {Schwarz}, {Visser}, \& {Zhang}}]{Bergin2016}
{Bergin}, E.~A., {Du}, F., {Cleeves}, L.~I., {et~al.} 2016, \apj, 831, 101

\bibitem[{{Bethell} \& {Bergin}(2009)}]{Bethell2009}
{Bethell}, T. \& {Bergin}, E. 2009, Science, 326, 1675

\bibitem[{{Carr} \& {Najita}(2011)}]{Carr2011}
{Carr}, J.~S. \& {Najita}, J.~R. 2011, \apj, 733, 102

\bibitem[{{Cazaux} \& {Tielens}(2004)}]{Cazaux2004}
{Cazaux}, S. \& {Tielens}, A.~G.~G.~M. 2004, \apj, 604, 222

\bibitem[{{Cazaux} \& {Tielens}(2010)}]{Cazaux2010}
{Cazaux}, S. \& {Tielens}, A.~G.~G.~M. 2010, \apj, 715, 698

\bibitem[{{Cherchneff} \& {Glassgold}(1993)}]{Cherchneff1993}
{Cherchneff}, I. \& {Glassgold}, A.~E. 1993, \apjl, 419, L41

\bibitem[{{Cherchneff} {et~al.}(1993){Cherchneff}, {Glassgold}, \& {Mamon}}]{1993ApJ...410..188C}
{Cherchneff}, I., {Glassgold}, A.~E., \& {Mamon}, G.~A. 1993, \apj, 410, 188

\bibitem[{{Draine} \& {Bertoldi}(1996)}]{Draine1996}
{Draine}, B.~T. \& {Bertoldi}, F. 1996, \apj, 468, 269

\bibitem[{{Duval} {et~al.}(2022){Duval}, {Bosman}, \& {Bergin}}]{Duval2022...934L..25D}
{Duval}, S.~E., {Bosman}, A.~D., \& {Bergin}, E.~A. 2022, \apjl, 934, L25

\bibitem[{Greenwood {et~al.}(2019{\natexlab{a}})Greenwood, Kamp, Waters, Woitke, \& Thi}]{Greenwood2019}
Greenwood, A.~J., Kamp, I., Waters, L. B. F.~M., Woitke, P., \& Thi, W.~F. 2019{\natexlab{a}}, Astronomy and Astrophysics, 626, A6

\bibitem[{Greenwood {et~al.}(2019{\natexlab{b}})Greenwood, Kamp, Waters, Woitke, \& Thi}]{Greenwood2019a}
Greenwood, A.~J., Kamp, I., Waters, L. B. F.~M., Woitke, P., \& Thi, W.~F. 2019{\natexlab{b}}, Astronomy and Astrophysics, 631, A81

\bibitem[{{Guzm{\'a}n} {et~al.}(2021){Guzm{\'a}n}, {Bergner}, {Law}, {{\"O}berg}, {Walsh}, {Cataldi}, {Aikawa}, {Bergin}, {Czekala}, {Huang}, {Andrews}, {Loomis}, {Zhang}, {Le Gal}, {Alarc{\'o}n}, {Ilee}, {Teague}, {Cleeves}, {Wilner}, {Long}, {Schwarz}, {Bosman}, {P{\'e}rez}, {M{\'e}nard}, \& {Liu}}]{viv2021}
{Guzm{\'a}n}, V.~V., {Bergner}, J.~B., {Law}, C.~J., {et~al.} 2021, \apjs, 257, 6

\bibitem[{{Heays} {et~al.}(2017){Heays}, {Bosman}, \& {van Dishoeck}}]{Heays2017}
{Heays}, A.~N., {Bosman}, A.~D., \& {van Dishoeck}, E.~F. 2017, \aap, 602, A105

\bibitem[{{Henning} \& {Semenov}(2013)}]{Henning2013}
{Henning}, T. \& {Semenov}, D. 2013, Chemical Reviews, 113, 9016

\bibitem[{Hörst(2017)}]{horst2017}
Hörst, S.~M. 2017, Journal of Geophysical Research: Planets, 122, 432

\bibitem[{{Ilee} {et~al.}(2021){Ilee}, {Walsh}, {Booth}, {Aikawa}, {Andrews}, {Bae}, {Bergin}, {Bergner}, {Bosman}, {Cataldi}, {Cleeves}, {Czekala}, {Guzm{\'a}n}, {Huang}, {Law}, {Le Gal}, {Loomis}, {M{\'e}nard}, {Nomura}, {{\"O}berg}, {Qi}, {Schwarz}, {Teague}, {Tsukagoshi}, {Wilner}, {Yamato}, \& {Zhang}}]{JD2021A}
{Ilee}, J.~D., {Walsh}, C., {Booth}, A.~S., {et~al.} 2021, \apjs, 257, 9

\bibitem[{{Kalv{\={a}}ns}(2021)}]{Kalvans2021}
{Kalv{\={a}}ns}, J. 2021, \apj, 910, 54

\bibitem[{Kamp {et~al.}(2017)Kamp, Thi, Woitke, Rab, Bouma, \& M{\'{e}}nard}]{Kamp2017}
Kamp, I., Thi, W.~F., Woitke, P., {et~al.} 2017, Astronomy and Astrophysics, 607, A41

\bibitem[{{Kamp} {et~al.}(2010){Kamp}, {Tilling}, {Woitke}, {Thi}, \& {Hogerheijde}}]{Kamp2010}
{Kamp}, I., {Tilling}, I., {Woitke}, P., {Thi}, W.~F., \& {Hogerheijde}, M. 2010, \aap, 510, A18

\bibitem[{Kress {et~al.}(2010)Kress, Tielens, \& Frenklach}]{Kress2010}
Kress, M.~E., Tielens, A. G. G.~M., \& Frenklach, M. 2010, Advances in Space Research, 46, 44

\bibitem[{{Loison} {et~al.}(2019){Loison}, {Dobrijevic}, \& {Hickson}}]{Loison2019}
{Loison}, J.~C., {Dobrijevic}, M., \& {Hickson}, K.~M. 2019, \icarus, 329, 55

\bibitem[{{McElroy} {et~al.}(2013){McElroy}, {Walsh}, {Markwick}, {Cordiner}, {Smith}, \& {Millar}}]{McElroy2013}
{McElroy}, D., {Walsh}, C., {Markwick}, A.~J., {et~al.} 2013, \aap, 550, A36

\bibitem[{{McEwan} {et~al.}(1999){McEwan}, {Scott}, {Adams}, {Babcock}, {Terzieva}, \& {Herbst}}]{McEwan1999}
{McEwan}, M.~J., {Scott}, G. B.~I., {Adams}, N.~G., {et~al.} 1999, \apj, 513, 287

\bibitem[{{Meijerink} {et~al.}(2012){Meijerink}, {Aresu}, {Kamp}, {Spaans}, {Thi}, \& {Woitke}}]{Meijerink2012}
{Meijerink}, R., {Aresu}, G., {Kamp}, I., {et~al.} 2012, \aap, 547, A68

\bibitem[{{Millar} \& {Herbst}(1994)}]{Millar1994}
{Millar}, T.~J. \& {Herbst}, E. 1994, \aap, 288, 561

\bibitem[{{Pontoppidan} {et~al.}(2010){Pontoppidan}, {Salyk}, {Blake}, {Meijerink}, {Carr}, \& {Najita}}]{Pontoppidan2010}
{Pontoppidan}, K.~M., {Salyk}, C., {Blake}, G.~A., {et~al.} 2010, \apj, 720, 887

\bibitem[{{Qi} {et~al.}(2013){Qi}, {{\"O}berg}, {Wilner}, \& {Rosenfeld}}]{Qi2013a}
{Qi}, C., {{\"O}berg}, K.~I., {Wilner}, D.~J., \& {Rosenfeld}, K.~A. 2013, \apjl, 765, L14

\bibitem[{{Rab} {et~al.}(2018){Rab}, {G{\"u}del}, {Woitke}, {Kamp}, {Thi}, {Min}, {Aresu}, \& {Meijerink}}]{Crab2018}
{Rab}, C., {G{\"u}del}, M., {Woitke}, P., {et~al.} 2018, \aap, 609, A91

\bibitem[{Rimmer \& Helling(2016)}]{Rimmer2016}
Rimmer, P.~B. \& Helling, C. 2016, The Astrophysical Journal Supplement Series, 224, 9

\bibitem[{{Rothman} {et~al.}(2009){Rothman}, {Gordon}, {Barbe}, {Benner}, {Bernath}, {Birk}, {Boudon}, {Brown}, {Campargue}, {Champion}, {Chance}, {Coudert}, {Dana}, {Devi}, {Fally}, {Flaud}, {Gamache}, {Goldman}, {Jacquemart}, {Kleiner}, {Lacome}, {Lafferty}, {Mandin}, {Massie}, {Mikhailenko}, {Miller}, {Moazzen-Ahmadi}, {Naumenko}, {Nikitin}, {Orphal}, {Perevalov}, {Perrin}, {Predoi-Cross}, {Rinsland}, {Rotger}, {{\v{S}}ime{\v{c}}kov{\'a}}, {Smith}, {Sung}, {Tashkun}, {Tennyson}, {Toth}, {Vandaele}, \& {Vander Auwera}}]{Rothman2009}
{Rothman}, L.~S., {Gordon}, I.~E., {Barbe}, A., {et~al.} 2009, \jqsrt, 110, 533

\bibitem[{{Sakai} {et~al.}(2008){Sakai}, {Sakai}, {Hirota}, \& {Yamamoto}}]{Sakai2008}
{Sakai}, N., {Sakai}, T., {Hirota}, T., \& {Yamamoto}, S. 2008, \apj, 672, 371

\bibitem[{Sakai \& Yamamoto(2013)}]{Sakai2013}
Sakai, N. \& Yamamoto, S. 2013, Chemical Reviews, 113, 8981

\bibitem[{{Salyk} {et~al.}(2008){Salyk}, {Pontoppidan}, {Blake}, {Lahuis}, {van Dishoeck}, \& {Evans}}]{Salyk2008}
{Salyk}, C., {Pontoppidan}, K.~M., {Blake}, G.~A., {et~al.} 2008, \apjl, 676, L49

\bibitem[{Santoro {et~al.}(2020)Santoro, Mart{\'{i}}nez, Lauwaet, Accolla, Tajuelo-Castilla, Merino, Sobrado, Pel{\'{a}}ez, Herrero, Tanarro, Mayoral, Ag{\'{u}}ndez, Sabbah, Joblin, Cernicharo, \& Mart{\'{i}}n-Gago}]{Santoro2020}
Santoro, G., Mart{\'{i}}nez, L., Lauwaet, K., {et~al.} 2020, The Astrophysical Journal, 895, 97

\bibitem[{{Tabone} {et~al.}(2023){Tabone}, {Bettoni}, {van Dishoeck}, {Arabhavi}, {Grant}, {Gasman}, {Henning}, {Kamp}, {G{\"u}del}, {Lagage}, {Ray}, {Vandenbussche}, {Abergel}, {Absil}, {Argyriou}, {Barrado}, {Boccaletti}, {Bouwman}, {Caratti o Garatti}, {Geers}, {Glauser}, {Justannont}, {Lahuis}, {Mueller}, {Nehm{\'e}}, {Olofsson}, {Pantin}, {Scheithauer}, {Waelkens}, {Waters}, {Black}, {Christiaens}, {Guadarrama}, {Morales-Calder{\'o}n}, {Jang}, {Kanwar}, {Pawellek}, {Perotti}, {Perrin}, {Rodgers-Lee}, {Samland}, {Schreiber}, {Schwarz}, {Colina}, {{\"O}stlin}, \& {Wright}}]{Tabone2023}
{Tabone}, B., {Bettoni}, G., {van Dishoeck}, E.~F., {et~al.} 2023, Nature Astronomy [\eprint[arXiv]{2304.05954}]

\bibitem[{{Vuitton} {et~al.}(2019){Vuitton}, {Yelle}, {Klippenstein}, {H{\"o}rst}, \& {Lavvas}}]{Vuitton2019}
{Vuitton}, V., {Yelle}, R.~V., {Klippenstein}, S.~J., {H{\"o}rst}, S.~M., \& {Lavvas}, P. 2019, \icarus, 324, 120

\bibitem[{{Wakelam} {et~al.}(2012){Wakelam}, {Herbst}, {Loison}, {Smith}, {Chandrasekaran}, {Pavone}, {Adams}, {Bacchus-Montabonel}, {Bergeat}, {B{\'e}roff}, {Bierbaum}, {Chabot}, {Dalgarno}, {van Dishoeck}, {Faure}, {Geppert}, {Gerlich}, {Galli}, {H{\'e}brard}, {Hersant}, {Hickson}, {Honvault}, {Klippenstein}, {Le Picard}, {Nyman}, {Pernot}, {Schlemmer}, {Selsis}, {Sims}, {Talbi}, {Tennyson}, {Troe}, {Wester}, \& {Wiesenfeld}}]{Wakelam2012}
{Wakelam}, V., {Herbst}, E., {Loison}, J.~C., {et~al.} 2012, \apjs, 199, 21

\bibitem[{{Wakelam} {et~al.}(2015){Wakelam}, {Loison}, {Herbst}, {Pavone}, {Bergeat}, {B{\'e}roff}, {Chabot}, {Faure}, {Galli}, {Geppert}, {Gerlich}, {Gratier}, {Harada}, {Hickson}, {Honvault}, {Klippenstein}, {Le Picard}, {Nyman}, {Ruaud}, {Schlemmer}, {Sims}, {Talbi}, {Tennyson}, \& {Wester}}]{wakelam2015}
{Wakelam}, V., {Loison}, J.~C., {Herbst}, E., {et~al.} 2015, \apjs, 217, 20

\bibitem[{{Walsh} {et~al.}(2015){Walsh}, {Nomura}, \& {van Dishoeck}}]{Walsh2015}
{Walsh}, C., {Nomura}, H., \& {van Dishoeck}, E. 2015, \aap, 582, A88

\bibitem[{Woitke {et~al.}(2009)Woitke, Kamp, \& Thi}]{Woitke2009}
Woitke, P., Kamp, I., \& Thi, W.~F. 2009, Astronomy and Astrophysics, 501, 383

\bibitem[{Woitke {et~al.}(2016)Woitke, Min, Pinte, Thi, Kamp, Rab, Anthonioz, Antonellini, Baldovin-Saavedra, Carmona, Dominik, Dionatos, Greaves, G{\"{u}}del, Ilee, Liebhart, M{\'{e}}nard, Rigon, Waters, Aresu, Meijerink, \& Spaans}]{Woitke2016}
Woitke, P., Min, M., Pinte, C., {et~al.} 2016, Astronomy and Astrophysics, 586, A103

\bibitem[{{Woitke} {et~al.}(2018){Woitke}, {Min}, {Thi}, {Roberts}, {Carmona}, {Kamp}, {M{\'e}nard}, \& {Pinte}}]{Woitke2018}
{Woitke}, P., {Min}, M., {Thi}, W.~F., {et~al.} 2018, \aap, 618, A57

\bibitem[{Woods \& Willacy(2007)}]{Woods2007}
Woods, P.~M. \& Willacy, K. 2007, The Astrophysical Journal, 655, L49

\end{thebibliography}

\appendix
\onecolumn
\section{Reactions taken from STAND2020}\label{3body reactions}
We list the three body and thermal decomposition reactions that are taken from STAND2020 network and are included in our models. As three body reactions are pressure dependent, we take the low pressure rate coefficients for these reactions. A detailed analysis to formulate a complete three body reaction network will be left for future work.

\begin{table*}[]
\caption{The list of three body and thermal decomposition reactions added.}
\label{three body reactions}
\begin{tabular}{llll}
\hline
Reactions                                                & $\alpha$ & $\beta$   & $\gamma$  \\ \hline
CH + M $\rightarrow$ C + H + M                           & 3.16E-10 & 0.00E+00  & 3.37E+04  \\ 
CN + M $\rightarrow$ C + N + M                           & 1.09E-9  & 0.00E+00  & 7.10E+04  \\
H + H + M $\rightarrow$ H$_2$ + M                        & 9.13E-33 & -6.00E-01 & 0.00E+00  \\
H + N + M $\rightarrow$ NH + M                             & 5.02E-32 & 0.00E+00  & 0.00E+00  \\
N + N + M $\rightarrow$ N$_2$ + M                          & 1.25E-32 & 0.00E+00  & 0.00E+00  \\
N + O + M $\rightarrow$ NO + M                             & 3.26E-33 & 0.00E+00  & 0.00E+00  \\
C$_2$H + M $\rightarrow$ C$_2$ + H + M                     & 5.00E-01 & -5.16E+00 & 5.74E+04  \\
C$_2$O + M $\rightarrow$ C$_2$ + O + M                     & 5.00E-01 & -5.16E+00 & 5.74E+04  \\
CH$_2$ + M $\rightarrow$ CH + H + M                        & 9.33E+00 & 0.00E+00  & 4.51E+04  \\
HCN +  M $\rightarrow$ HNC +  M                            & 1.45E-06 & 1.00E+00  & 2.38E+04  \\
CN + H + M $\rightarrow$ HCN + M                           & 9.35E-30 & -2.00E+00 & 5.20E+20  \\
CO + O + M $\rightarrow$ CO$_2$ + M                        & 1.20E-32 & 0.00E+00  & 2.16E+03  \\
NH$_2$ + M $\rightarrow$ NH + H + M                        & 1.99E-09 & 0.00E+00  & 3.83E+04  \\
H + CH$_2$ + M $\rightarrow$ CH$_3$+ M                     & 5.63E-31 & 0.00E+00  & 0.00E+00  \\
NH$_3$ + M $\rightarrow$ NH + H$_2$ + M                    & 1.05E-09 & 0.00E+00  & 4.70E+04  \\
OCN + M $\rightarrow$ CO + N + M                           & 3.95E-06 & -1.90E+00 & 3.01E+04  \\
H + C$_2$H + M $\rightarrow$ C$_2$H$_2$ + M                & 2.63E-26 & -3.10E+00 & 7.21E+02  \\
\ce{H2CO} + M $\rightarrow$ CO + \ce{H2} + M               & 9.40E-09 & 0.00E+00  & 3.32E+04  \\
\ce{C2H2} + M + H $\rightarrow$ M + \ce{C2H3} + M          & 4.87E-30 & -1.07E+00 & 8.38E+01   \\
\ce{H} + \ce{C2H4} + M $\rightarrow$ \ce{C2H5} + M         & 9.23E-29 & -1.51E+00 & 7.29E+01  \\
\ce{H} + \ce{C2H5} + M $\rightarrow$ \ce{CH3CH3} + M       & 2.00E-28 & -1.50E+00 & 0.00E+00  \\
H + \ce{C2H3} + M $\rightarrow$ \ce{C2H4} + M              & 1.49E-29 & -1.00E+00 & 0.00E+00   \\
\ce{CH2OH} + M + H $\rightarrow$ \ce{CH3OH} + M            & 1.20E-29 & 1.04E+00  & 0.00E+00  \\
\ce{CH3O} + H + M $\rightarrow$ \ce{CH3OH} + M             & 7.21E-30 & 1.24E+00  & 0.00E+00  \\
\ce{H2CO} + H + M $\rightarrow$ \ce{CH3O} + M              & 1.80E-31 & 6.60E-01  & 8.63E+02  \\
\ce{CH2OH} +  M $\rightarrow$ \ce{H2CO} + H + M            & 1.66E-10 & 0.00E+00  & 1.26E+04  \\
\ce{C2H} + M $\rightarrow$ \ce{C2} + H + M                 & 5.00E-01 & -5.16E+00 & 5.74E+04  \\
\ce{C2O} + M $\rightarrow$ \ce{C2} + O + M                 & 5.00E-01 & -5.16E+00 & 5.74E+04  \\
H + \ce{C2H3} + M $\rightarrow$ \ce{C2H4} + M              & 1.49E-29 & -1.00E+00 & 0.00E+00  \\
\ce{C2H4} +  M $\rightarrow$ \ce{C2H2} + \ce{H2} + M       & 5.80E-08 & 0.00E+00  & 3.60E+00   \\
H + \ce{C4H} + M $\rightarrow$ \ce{HC4H} + M               & 2.64E-26 & -3.10E+00 & 7.21E+02  \\
\ce{C2H} + \ce{C2H} + M $\rightarrow$ \ce{HC4H} + M        & 5.56E-28 & -3.00E+00 & 3.00E+02  \\
H + \ce{C4H3} +  M $\rightarrow$ \ce{CH2CHCCH} + M         & 5.41E-23 & -3.97E+00 & 1.77E+02  \\
H + \ce{HC4H} + M $\rightarrow$ \ce{C4H3} + M              & 3.53E-25 & -2.93E+00 & 1.76E+02  \\
\ce{C2H2} + \ce{C2H2} +  M $\rightarrow$ \ce{CH2CHCCH} + M & 3.78E-31 & 1.00E+00  & 1.86E+00   \\
\ce{CH3CH3} +  M $\rightarrow$ \ce{C2H4} +  \ce{H2} + M    & 3.80E-07 & 0.00E+00  & 3.40E+04  \\
\ce{CH2CCH} + M $\rightarrow$ \ce{C2H2} +  CH + M          & 1.00E-14 & 0.00E+00  & 0.00E+00  \\
\ce{C4H3} + M $\rightarrow$ \ce{C2H2} +  \ce{C2H} +  M     & 5.34E-05 & 1.00E+00  & 4.15E+04  \\
H + \ce{e-} + M $\rightarrow$ \ce{H-} + M                  & 2.50E-31 & -1.50E+00 & 0.00E+00  \\
HS + H + M $\rightarrow$ \ce{H2S}  + M                     & 1.00E-30 & -2.00E+00 & 0.00E+00  \\
\ce{H2} +  S + M $\rightarrow$ \ce{H2S} + M                & 1.40E-31 & -1.90E+00 & 8.14E+03  \\
CO + S + M $\rightarrow$ OCS + M                           & 3.00E-33 & 0.00E+00  & 1.00E+03  \\
\ce{CH3+} +  \ce{H2} + M $\rightarrow$ \ce{CH5+} +  M      & 1.10E-28 & 0.00E+00  & 0.00E+00  \\
\ce{C2H2+} + \ce{C2H2} + M $\rightarrow$ \ce{C4H4+} + M    & 1.60E-26 & 0.00E+00  & 0.00E+00  \\
\ce{C2H2+} + \ce{H2} + M $\rightarrow$ \ce{C2H4+} + M      & 1.20E-27 & 0.00E+00  & 0.00E+00  \\
\ce{C2H3+} + \ce{H2} + M $\rightarrow$ \ce{C2H5+} + M      & 1.49E-29 & 0.00E+00  & 0.00E+00 \\ \hline
\end{tabular}%
\end{table*}

\section{\ce{C2H2} formation/destruction pathways in KIDA2014}
The following figures show the major formation and destruction pathways leading to \ce{C2H2} when using the KIDA2014 database in model 2 (using large DIANA network) and model 4 (using extended hydrocarbon network).

\begin{figure*}[h]
    \centering
    \includegraphics[width=.7\linewidth]{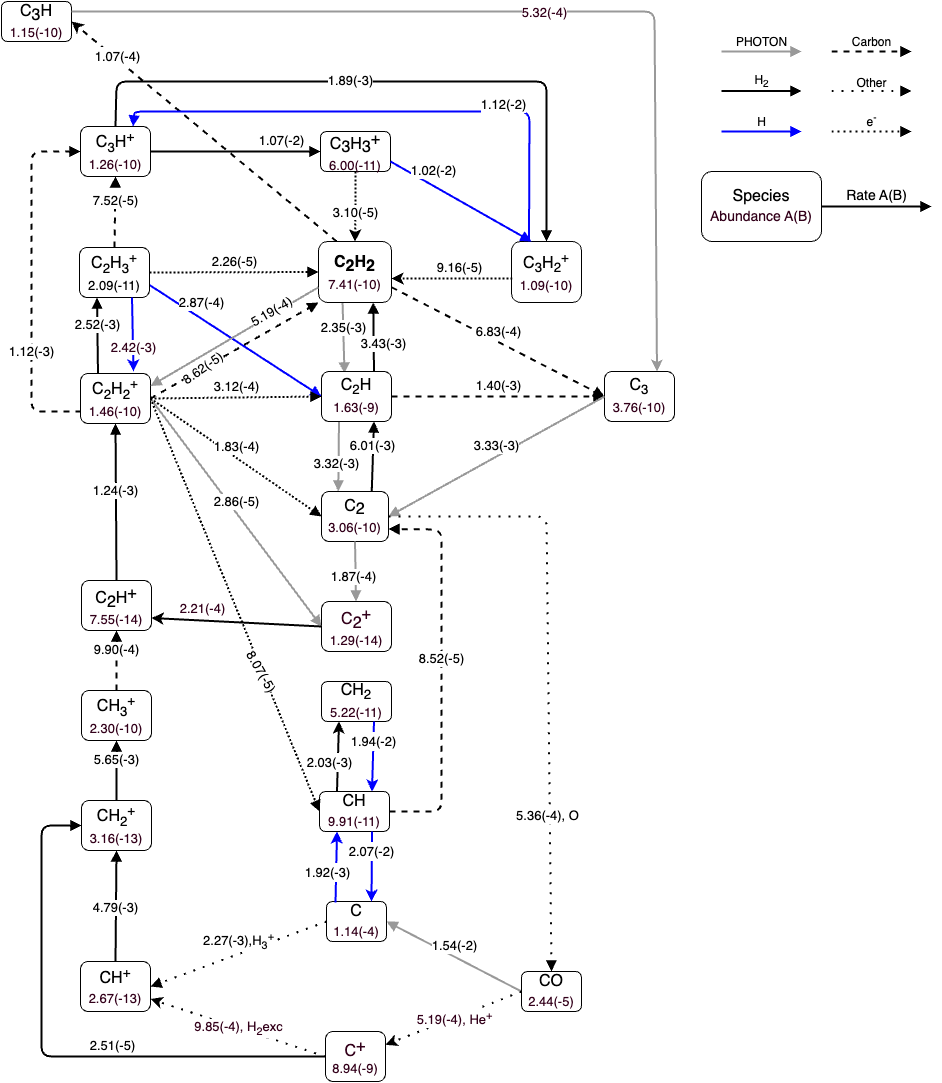}
    \caption{The chemical network centered around C$_2$H$_2$ depicting the dominant formation and destruction pathways of parent molecules for model~2 (large DIANA network using the KIDA2014 rate database).}
\label{KIDA+317a}
\end{figure*}
\begin{figure*}[b]
    \centering
    \includegraphics[width=0.7\linewidth]{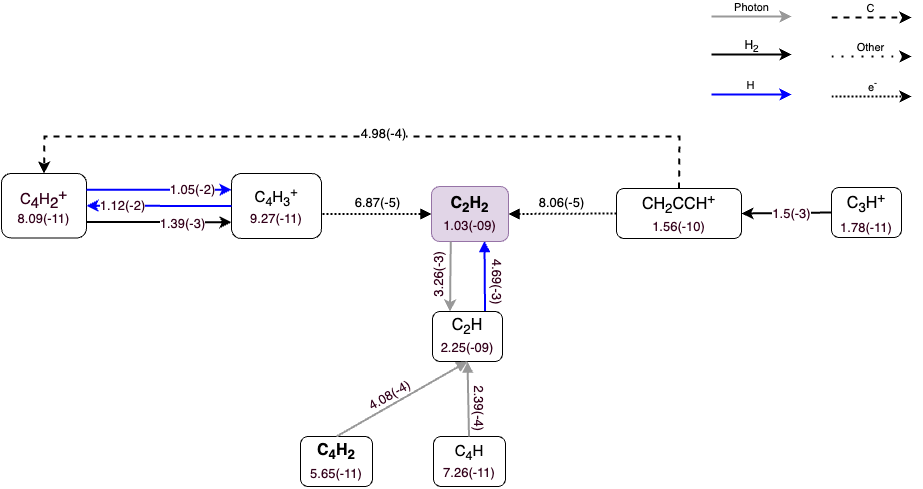}
    \caption{Zoomed in chemical network for the extended hydrocarbon chemistry in model~4 (KIDA2014) showing the new pathways forming \ce{C2H2} and \ce{C2H} that were not dominant in model~2 (UMIST2012). Shown are only two steps in the formation pathway.}
    \label{KIDA+317}
\end{figure*}

\end{document}